\documentclass[a4paper,11pt]{article}
\pdfoutput=1 
\usepackage{jinstpub} 
 
\usepackage[section]{placeins}
\RequirePackage{graphicx}
\usepackage{lineno}
\usepackage{ulem}

\title{Development and Performance of a Sealed Liquid Xenon Time Projection Chamber}

\author{Yuehuan Wei,}
\author[1]{Jianyu Long\note{Current institution: Department of Physics, University of Chicago},}
\author[2]{Francesco Lombardi\note{Current institution: Department of Physics,  University of Coimbra},}
\author[]{Zhiheng Jiang,}
\author[]{Jingqiang Ye,}
\author[3]{Kaixuan~Ni\note{Corresponding author},}

\affiliation{Department of Physics, University of California San Diego, La Jolla, CA, 92093, USA}

\emailAdd{nikx@physics.ucsd.edu}

\abstract{The liquid xenon (LXe) time projection chamber (TPC) technology is probing a wide range of dark matter masses from sub-GeV to a few TeV. To further improve  its sensitivity to sub-GeV dark matter and its application in reactor neutrino monitoring via coherent elastic neutrino-nucleus scattering (CE$\nu$NS), more understanding and suppression of single/few electrons background rate are needed. Here we report on the design and performance of a sealed LXeTPC with a graphene-coated fused silica window as the cathode. The purpose of the sealed TPC is for isolating the liquid xenon target volume from the majority of out-gassing materials in the detector vessel, thus improving the liquid xenon purification efficiency and reducing the impurity-induced single/few electrons background. We investigated the out-gassing rate and purification efficiency using the data from the sealed TPC with a simple purification model. The single electron signals from the photoionization of impurities in LXe are obtained and their correlation with the LXe purity is investigated. The photo-electron emission rate on the graphene-coated electrode is compared to that from stainless steel, the electrode material typically used in LXe detectors. We discuss the possible further improvement and potential applications of the sealed TPC for the next generation liquid xenon experiments for dark matter and neutrino physics.}


\arxivnumber{2007.16194} 

\begin{document}
\maketitle
\flushbottom

\section{Introduction}
\label{sec:intro}

Liquid Xenon Time Projection Chamber (LXeTPC) technology is widely used in rare event searches, such as dark matter (DM) direct detection ~\cite{Aprile:2018dbl, Cui:2017nnn, Akerib:2016vxi} and neutrino experiments~\cite{Anton:2019wmi}. In the field of DM direct detection, LXeTPC-based detectors have reached world-leading sensitivity to DM in the form of weakly interacting massive particles (WIMPs) with masses above a few GeV~\cite{Aprile:2018dbl}. LXe and liquid argon (LAr) detectors have also been used to search for sub-GeV DM via their scattering on electrons~\cite{Aprile:2019xxb, Aprile:2019jmx, Agnes:2018oej}, where the experimental limits on the DM cross section are less stringent. Improving the sensitivity of LXeTPCs to sub-GeV DM requires both better understanding and further suppression of background at the single- and a few-electrons level. A low background, single-electron sensitive LXe detector also has potential applications to detect coherent elastic neutrino-nucleus scattering (CE$\nu$NS) from reactor or solar neutrinos~\cite{Lenardo:2019fcn}.

A LXeTPC detects both scintillation light and ionization charge that are produced in particle interactions in the liquid xenon. The scintillation photons are detected using photomultiplier tubes (PMTs) and referred to as the S1 signal. The ionized electrons are drifted to the top of the TPC and extracted into a thin layer of gaseous xenon, where electroluminescence is produced on account of a strong electric field ($\sim$\,10 kV/cm), called proportional scintillation light. This proportional scintillation signal is detected by the same PMTs and referred to as the S2 signal. The $z$ coordinate of the interaction can be extracted from the time difference between S1 and S2, while the ($x$, $y$) coordinates can be reconstructed from the light pattern of S2 signal on top PMT array. The S2/S1 ratio is used to discriminate between nuclear and electronic recoils for background reduction. 

Benefiting from the amplification of S2 signals, LXeTPCs can detect very low energy events at a few electrons level~\cite{Aprile:2019xxb}. However, the background rate at single/few electrons level is still the limiting factor to improve the sensitivity. Photoionization on metal surfaces of TPC material, such as electrodes, and impurities in LXe are believed to be the main background source at the single/few electrons level. 

In this paper, we present a \textit{sealed} TPC with the goal of studying and suppressing the single- to few-electron background. This sealed TPC includes a cylindrical acrylic field cage and fused silica windows. In an attempt to further suppress the electrode-induced single electrons, a graphene-coated electrode is used. 


The design and operation of the sealed TPC are described in Sec.\,\ref{sec:sealedtpc}, including a COMSOL\,\cite{comsol} simulation for the design optimization. In Sec.\,\ref{result_discussion} we investigate the purification efficiency and out-gassing rate based on the LXe purity evolution and a simple purification model, followed by a discussion on the rates and production mechanisms of single electrons and their correlation with the LXe purity and electrode materials. 


\section{Detector design and operation}
\label{sec:sealedtpc}

\subsection{Design of the sealed TPC}
\label{sec:tpcdesign}

\begin{figure}[hbt!]
\centering
\begin{minipage}{0.5\columnwidth}
\centering
\includegraphics[width=\columnwidth]{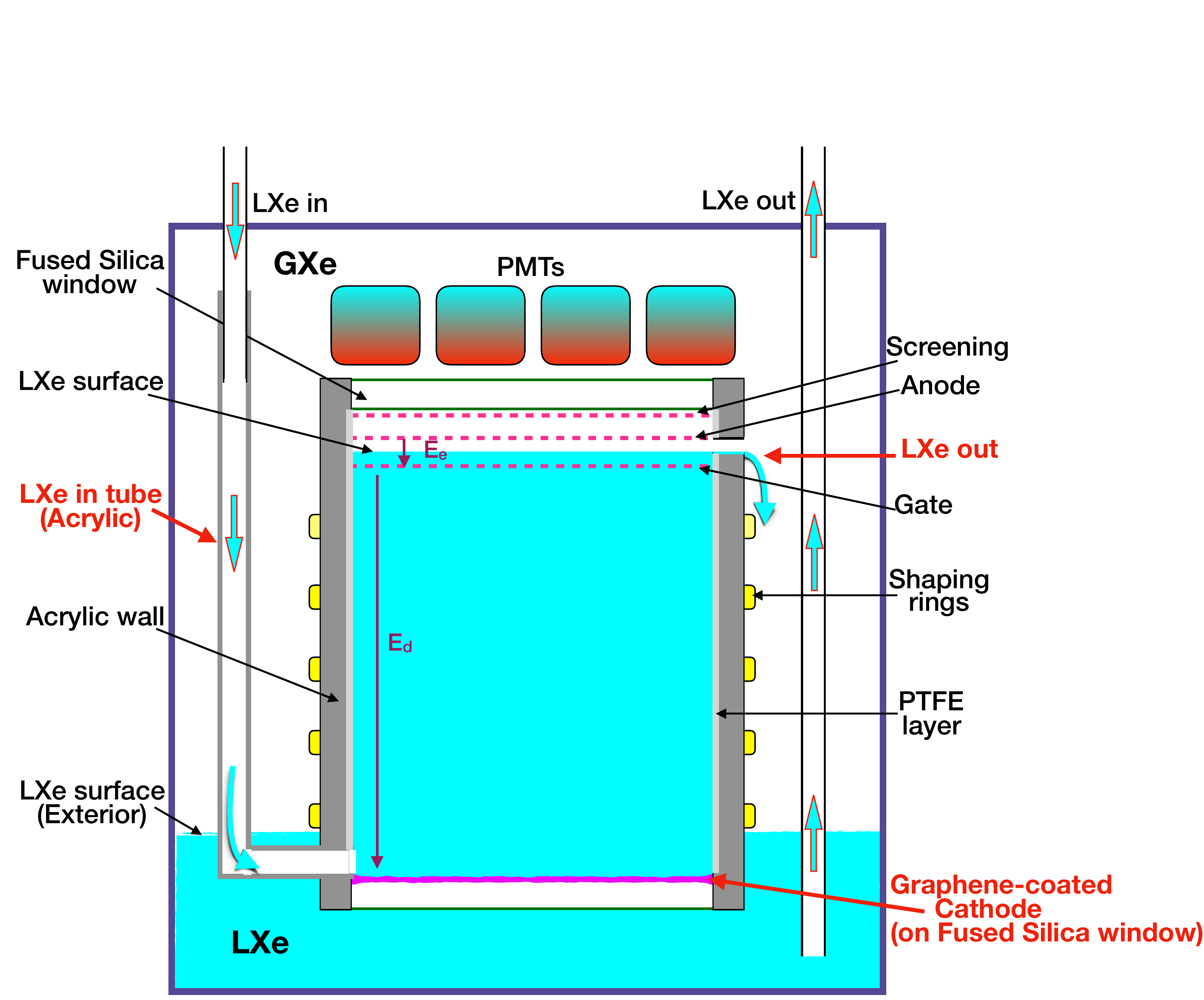}
\end{minipage}\hfill
\begin{minipage}{0.48\columnwidth}
\centering
\includegraphics[width=\columnwidth]{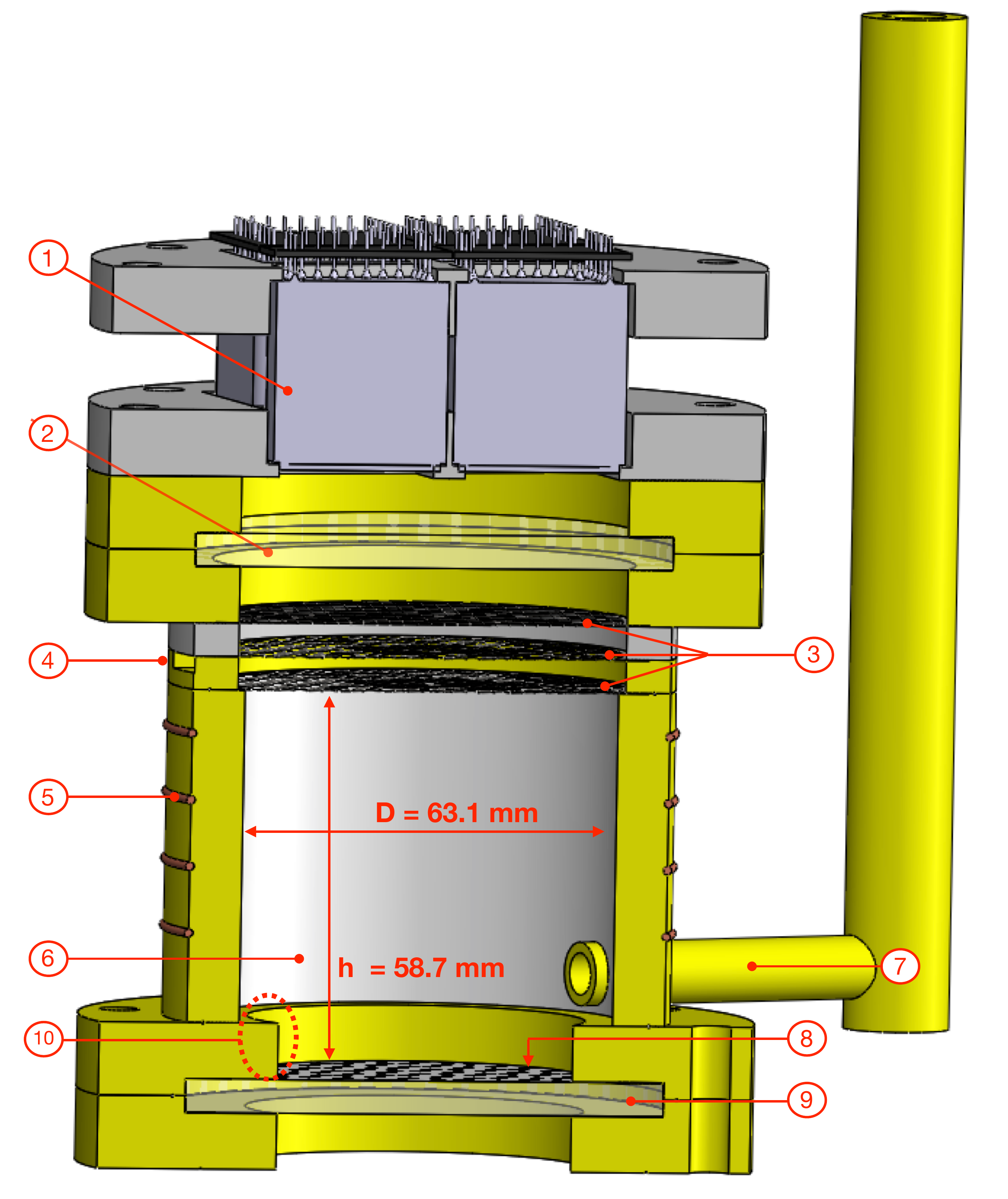}
\end{minipage}
\caption{(Left) The conceptual design of the sealed TPC. (Right) The 3D TPC design: 1 - Four 1"\,PMTs on top array. 2 - Fused silica window. 3 - TS, anode and gate electrodes. 4 - Slit for liquid level control. 5 - Field-shaping rings in copper. 6 - PTFE sheet for increasing light collection. 7 - Acrylic tube for liquid xenon filling. 8\,\&\,9 - Graphene cathode: graphene-coated on fused silica window. 10 - Protruding structure for holding the fused silica window. }
\label{fig:TPC_sketch}
\end{figure}

\begin{figure}[hbt!]
\begin{minipage}{0.44\columnwidth}
	\centering
	\includegraphics[width=\columnwidth]{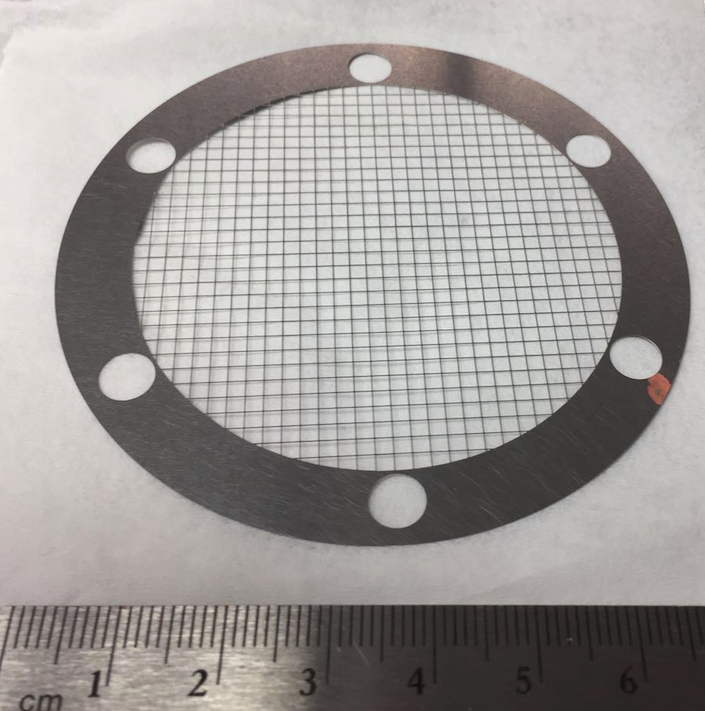}
    \end{minipage}\hfill
    \begin{minipage}{0.45\columnwidth}
	\centering
	\includegraphics[width=\columnwidth]{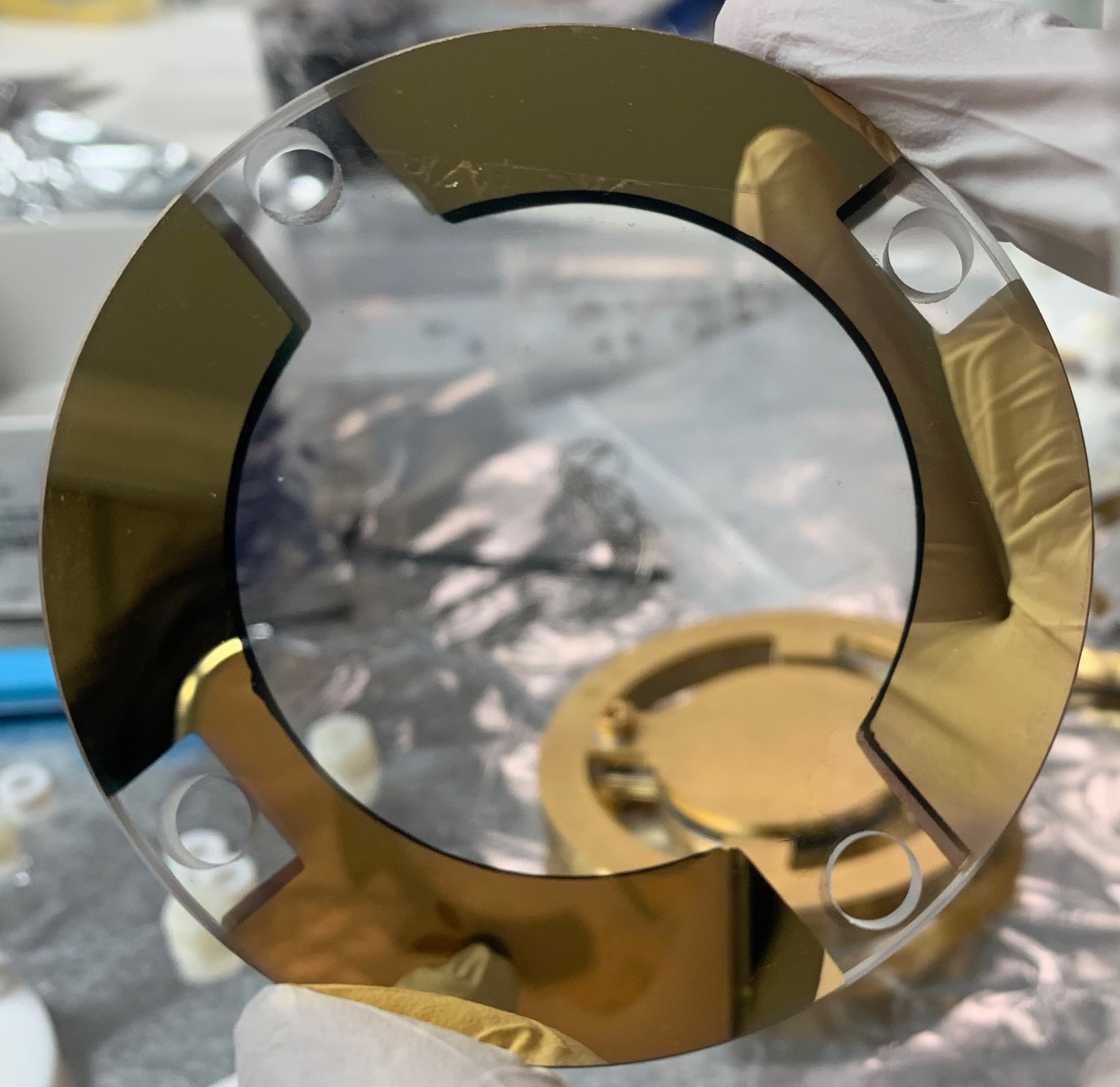}
    \end{minipage}
	\caption{(Left) Etched S.S mesh as TS, anode and gate electrodes. 
	         (Right) Graphene-coated fused silica window as cathode with gold-coated on the edge for HV connection.}
	\label{coating_pics}
\end{figure}

The schematic design of the sealed TPC is shown in figure\,\ref{fig:TPC_sketch} (left). A 3-dimensional (3D) mechanical design is shown on the right which reflects the actual size and structural details of each component. A target volume of $\sim\Phi$\,63\,mm $\times$ $h$\,59\,mm, corresponding to $0.54$\,kg of LXe, is defined by an acrylic cylinder. A grounding top screening (TS) electrode is installed to protect the PMTs from the high voltage (HV) from the anode electrode. Etched stainless steel (S.S) mesh, as shown in figure\,\ref{coating_pics} (left), is used as TS, anode and gate electrode respectively. The mesh pitch is 2\,mm and the wire diameter is 0.1\,mm, resulting in an optical transparency of 92\%. Two pieces of fused silica windows are placed at the top and bottom of the acrylic cylinder for isolating the LXe target volume from outside. Teflon o-rings are placed between the acrylic flange and fused silica windows for sealing. For the bottom fused silica window, a single-layer of graphene was coated on the surface to serve as the cathode electrode with $\sim$96\% transparency~\cite{Bae_2010}. The coating was done by Graphene Laboratories Inc.\,\cite{graphene_company}. To make a better HV connection, gold powder was coated on the edge of the fused silica window as shown in figure\,\ref{coating_pics} (right). Four field-shaping rings are connected via a resistor chain to ensure a uniform drift field between the negatively biased cathode and the gate at ground potential. The distance between the bottom shaping ring and cathode is doubled due to insufficient installation space caused by the protruding structure for holding the bottom silica window; thus, a resistor with twice the resistance value compared to the resistor chain above is used for this region. A thin layer of polytetrafluoroethylene\,(PTFE) is placed on the inner wall of the TPC to increase the light collection. Four 1"\,PMTs (Hamamatsu R8520) are placed above the top fused silica window for signal detection. 

The liquid surface is kept precisely in the middle of anode and gate electrodes at 2.5\,mm above gate, controlled by an open slit. During detector operation, the purified xenon is continuously circulated into the active volume of TPC through an acrylic tube. Once the liquid surface reaches the position of the slit, the overflowed LXe drips into the S.S vessel hosting the TPC, where it is then pumped out for purification by a SAES getter and re-condensed into the active target of TPC. This \textit{sealed} TPC design allows the purified xenon to be filled directly into the active LXe target without any contamination from the relatively less pure \textit{passive} liquid xenon outside of the field cage, resulting in an improved purification efficiency. The gas gap is given by the position of the liquid xenon overflow without the need of extra liquid level control.

\FloatBarrier

\FloatBarrier

\begin{figure}[hbt!]
\centering
\begin{minipage}{0.5\columnwidth}
\centering
\includegraphics[width=\columnwidth]{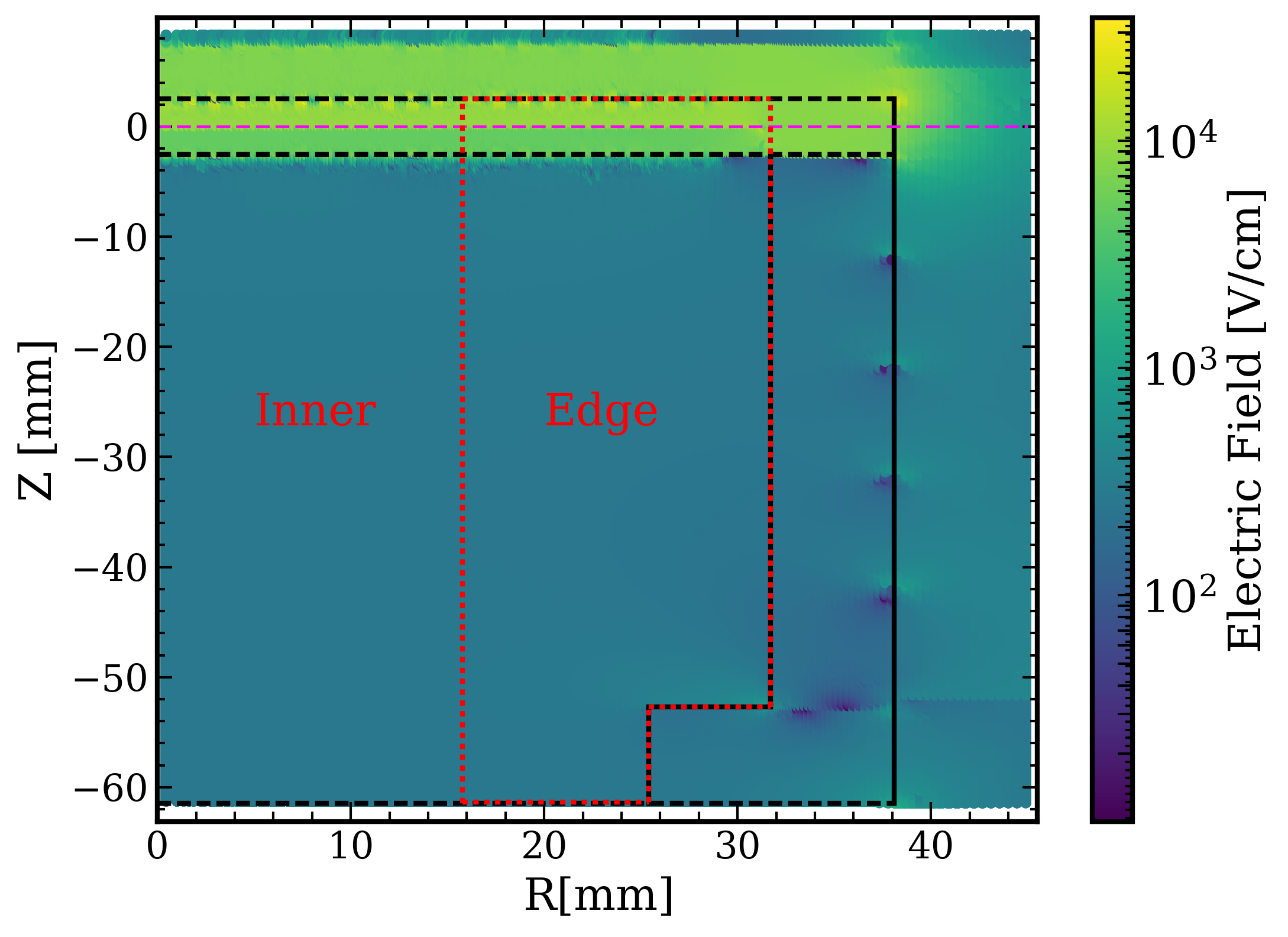}
\end{minipage}
\caption{The simulated electric field in TPC by \small{COMSOL} with anode and cathode at 4\,kV and -1.5\,kV, respectively.}
\label{electric_field}
\end{figure}

\begin{figure}[hbt!]
\begin{minipage}{0.5\columnwidth}
	\centering
	\includegraphics[width=\columnwidth]{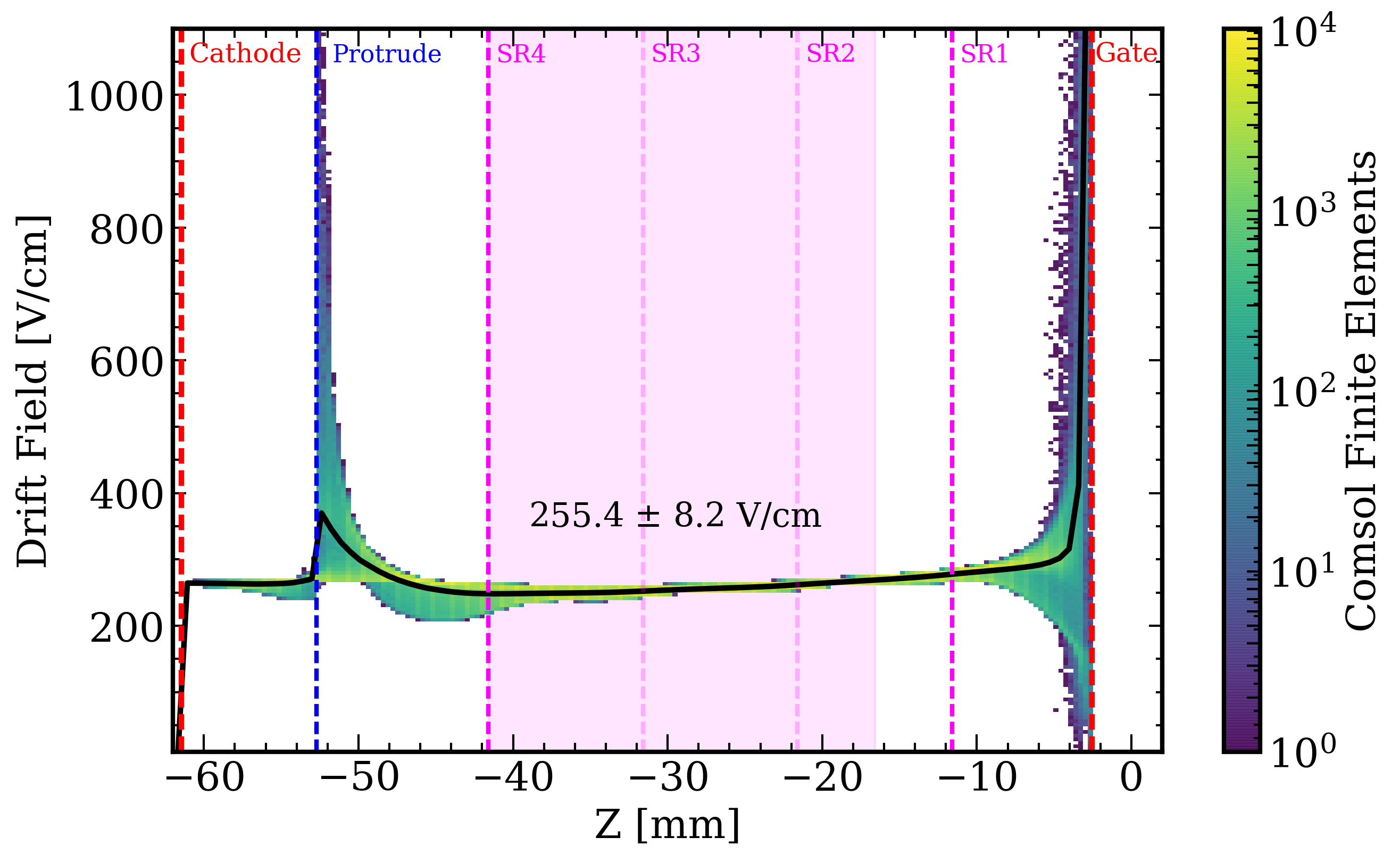}
    \end{minipage}\hfill
    \begin{minipage}{0.5\columnwidth}
	\centering
	\includegraphics[width=\columnwidth]{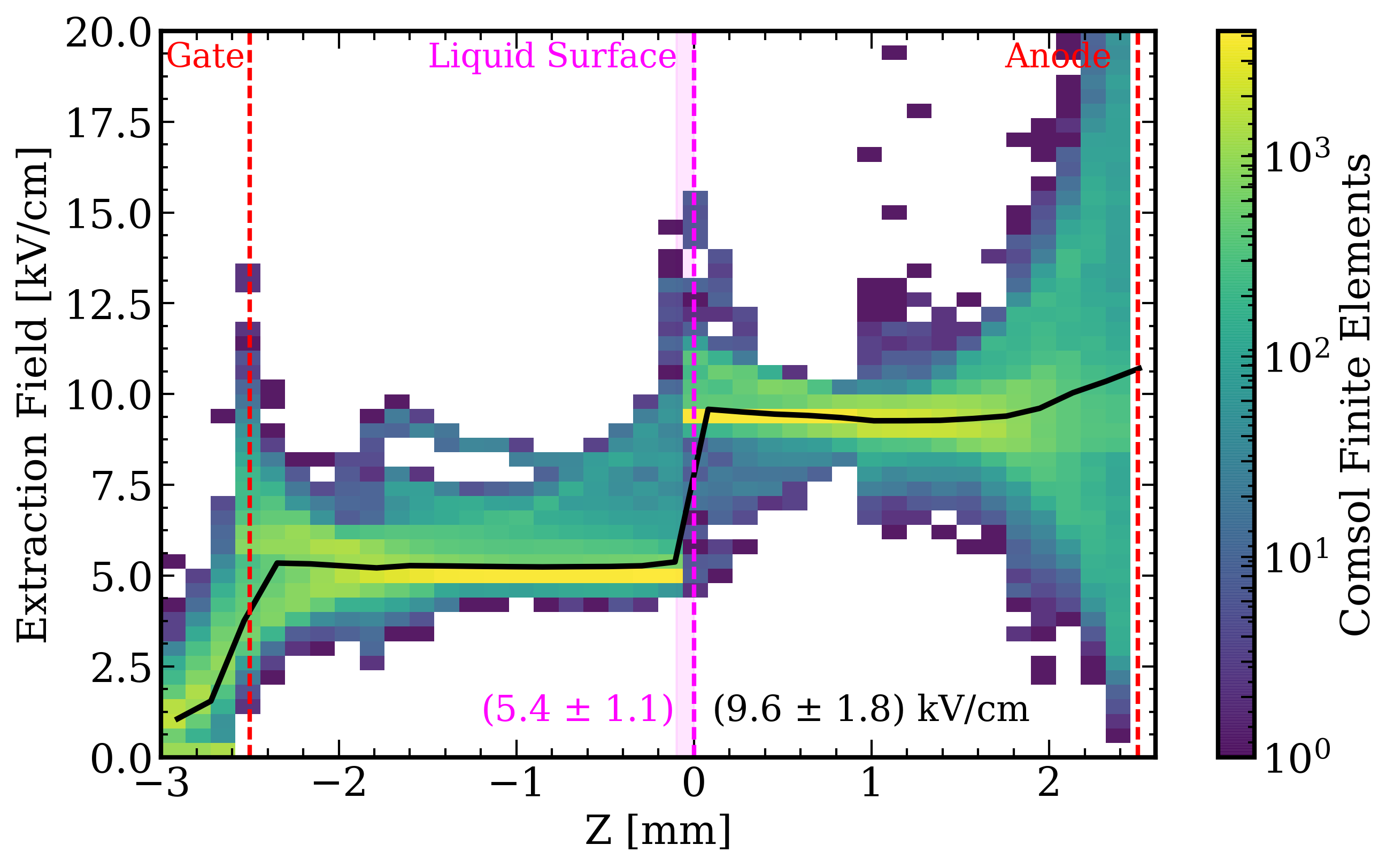}
    \end{minipage}
	\caption{Simulated drift field (left) and extraction field (right) as a function of $z$ in the edge region of figure\,\ref{electric_field}.}
	\label{extract_drift_field}
\end{figure}

To investigate the TPC field uniformity, an electric field simulation was performed using the finite element analysis software package COMSOL\,\cite{comsol}. A 3D model was built for precise estimation of field distribution. Limited by computing power, the etched meshes were modeled as parallel wires with diameter of 0.1\,mm, while the graphened-coated cathode was modeled as a plate. The error on the field calculation from the approximation is negligible compared with the variation of field in the LXe target. The field distribution from the simulation is shown in figure\,\ref{electric_field}. The anode and cathode are set at 4.0\,kV and -1.5\,kV respectively, which is the HV setting for all the presented data in the paper. The field distribution was investigated closely based on the events distribution from detector calibration. An external $^{57}$Co source was used for the calibration, as described in Sec.\,\ref{sec:co_cali}. Most of the events are localized in the edge region of the TPC due to the high self-shielding power of LXe as shown in figure\,\ref{fig:evts_dis} (left). The target volume of the TPC was divided into two parts, the center ($r$ < $\rm R$/2) and the edge ($r$ > $\rm R$/2), where $\rm R$ is the radius of the target volume. The drift and extraction fields in the edge region are shown in left and right of figure\,\ref{extract_drift_field}, respectively, as a function of $z$.  The black solid line is the mean field in each $z$ bin. The dashed-vertical lines in figure\,\ref{extract_drift_field}\,(left) represent the positions of the electrodes, shaping rings\,(SRs) and the protruding edge. The high field near the gate is caused by the field leaking from the high-field region between gate and anode, while the high field around the protruding edge is due to the dielectric constant difference between LXe and acrylic. To avoid the effect of field non-uniformity, the region from the middle of SR1 and SR2 to SR4 is used for the events selection in our main analysis, giving a drift field of (255$\,\pm\,$8)\,V/cm as shown in figure\,\ref{extract_drift_field}\,(left). The positions of the anode and gate electrodes and the liquid surface are marked as dashed-vertical lines in figure\,\ref{extract_drift_field}\,(right). A (5.4$\,\pm\,$1.1)\,kV/cm extraction field just below the liquid surface was achieved with the current HV setting. The field in the gas between the liquid surface and anode electrode is (9.6$\,\pm\,$1.8)\,kV/cm.

\FloatBarrier

\FloatBarrier

\subsection{Detector operation}
\label{sec:operation}

For stable operation of the TPC in a high purity LXe environment, a cryogenic system and a purification system was built. The cryogenic system consists of a pulse tube refrigerator (PTR) from Iwatani corporation, a heat exchanger and a vacuum insulated S.S vessel. The PTR consists of a PDC08 cold head and an SA115 helium compressor, providing 40\,W of cooling power at 165\,K. During operation liquefied xenon drips through a 1/4" S.S tube into the vessel where the TPC is located. Benefiting from the sealed structure, no liquid xenon is needed to fill the gap between the field cage and the wall of stainless steel vessel, requiring only a small amount of xenon for the active target inside the field cage.  In total 1.13\,kg xenon is used with a 0.54\,kg sensitive target inside the sealed TPC. The vessel is connected to a gas purification system, which contains a SAES Getter (model PS3-MT3-R-2) for removing impurities such as water and oxygen from xenon gas down to the part-per-billion (ppb) level. The detailed configuration of the cryogenic and purification system can be found in \cite{Lin:2013ypa}.

To investigate the detector performance, a 122\,keV  $^{57}$Co gamma ray source was placed externally for detector calibration and electron lifetime monitoring. The source was placed outside the detector at the same height as the TPC center during calibration. Without LXe surrounding the sealed TPC, the low energy gamma rays can easily reach the sensitive LXe target volume, thus producing an interaction rate sufficient for studying the detector performance. To record the signals, the PMT waveforms are digitized by CAEN V1720 FADC with a sampling frequency of 250\,MS/s after a low noise amplifier ($\times$10). The digital signals are read out by a PC via a CAEN A2818 optical controller and recorded for further offline analysis.

\section{Results and discussion}
\label{result_discussion}

\subsection{Calibration with $^{57}$Co}
\label{sec:co_cali}

\begin{figure}[hbt!]
    \centering
    \begin{minipage}{0.5\columnwidth}
        \includegraphics[width=\columnwidth]{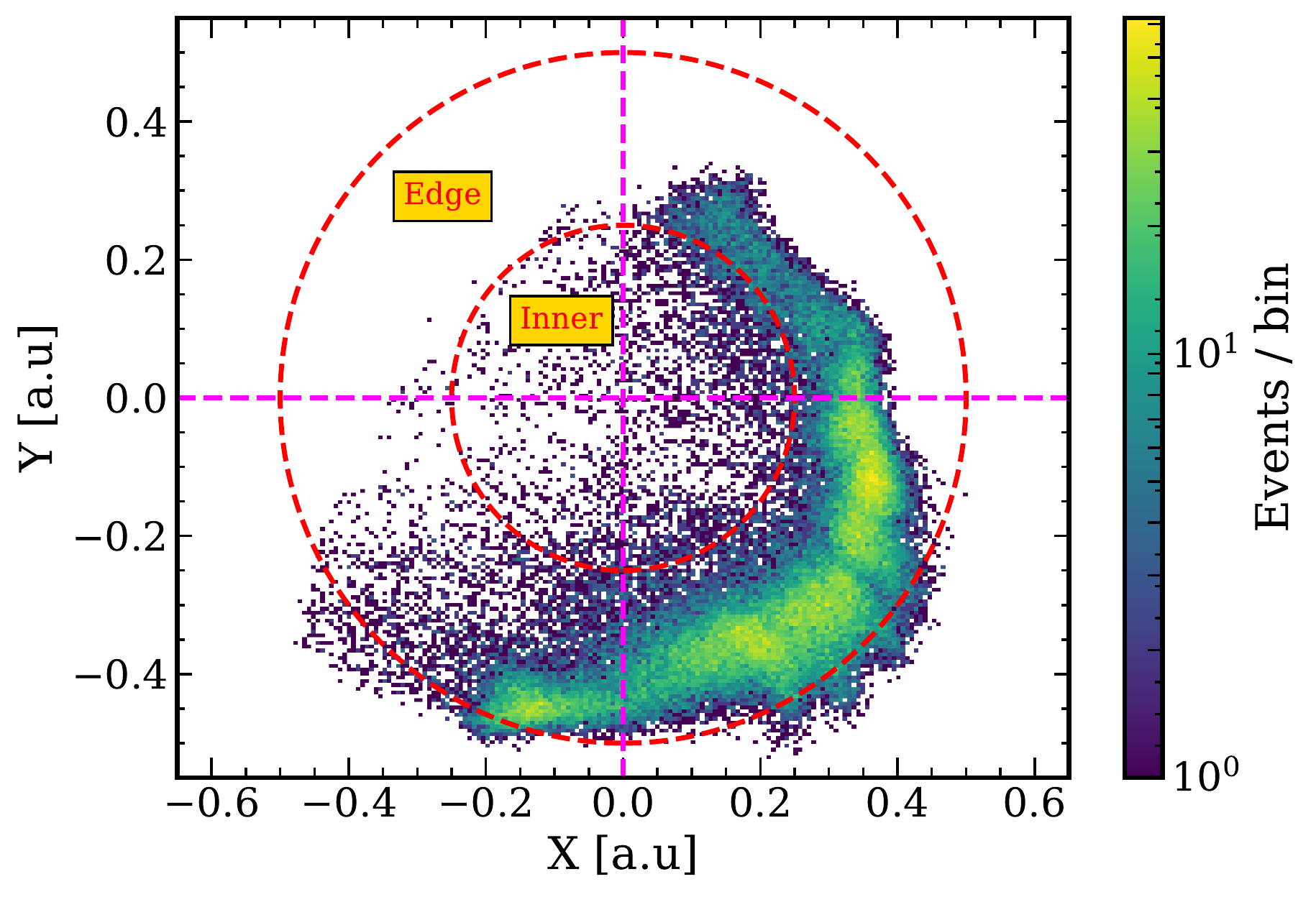}
    \end{minipage}\hfill
    \begin{minipage}{0.47\columnwidth}
        \includegraphics[width=\columnwidth]{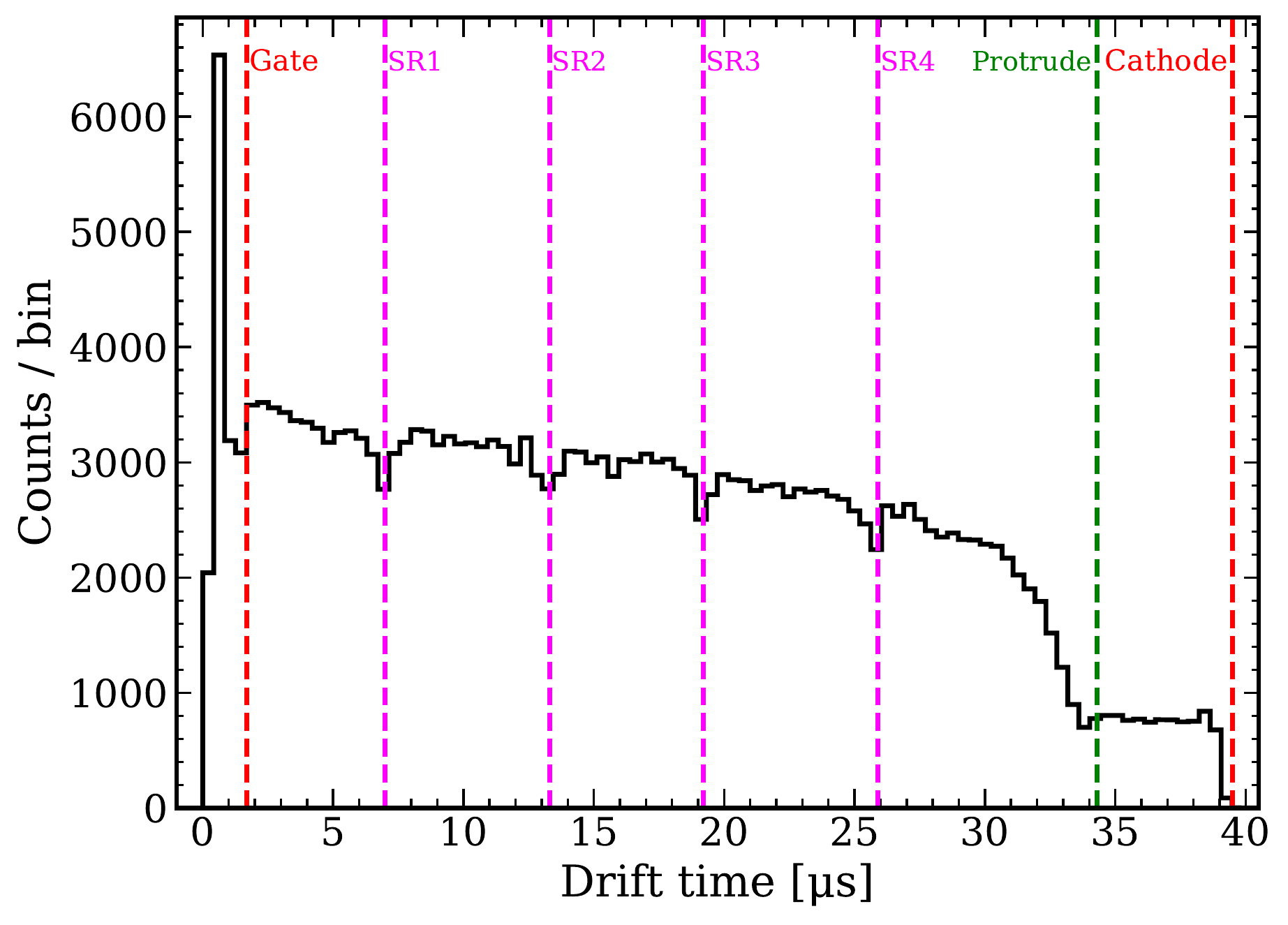}
    \end{minipage}\hfill
    \caption{(Left) Reconstructed ($x$, $y$) position based on a simple center-of-gravity method for $^{57}$Co  gamma rays interacting in the TPC. The red dashed circles denote the radius $R$ and $R$/2 of the LXe target volume respectively. 
    (Right) The drift time distribution. The red dashed lines indicate the location of gate and cathode electrodes respectively. The magenta dashed lines show the location of field-shaping rings while the green dashed line is the location of the protruding edge.}
    \label{fig:evts_dis}
\end{figure}

A two-phase xenon TPC allows for the reconstruction of 3D positions based on the detection of S1 and S2 signals. A simple ($x$, $y$) position reconstruction based on center-of-gravity method is used. Figure\,\ref{fig:evts_dis} (left) shows the reconstructed position of the events from an uncollimated $^{57}$Co source which is placed on the right-bottom in ($x$, $y$) coordinates. Most events are localized on the edge of the TPC, due to the strong self-shielding property of LXe. The position reconstruction capability is reduced for the edge events due to its insensitive to S2 pattern. The events in the top-right region that deviate from the TPC edge are mostly caused by the relatively low quantum efficiency of the particular PMT on that corner. The $z$ position can be inferred from the drift time of ionized electrons, i.e. the time difference between S1 and S2. In figure\,\ref{fig:evts_dis} (right), the location of the gate and cathode electrodes, denoted as red dashed lines, can be clearly observed in the drift time distribution. The locations of the four field-shaping rings are given by magenta dashed lines. The corresponding valleys are due to shielding of gamma rays from the copper rings. The descending step around $\sim$35\,$\mu$s is caused by the protruding structure as shown in figure\,\ref{fig:TPC_sketch} (right) for holding the bottom silica window. 

\FloatBarrier

\begin{figure}[hbt!]
    \centering
    \begin{minipage}{0.5\columnwidth}
        \includegraphics[width=\columnwidth]{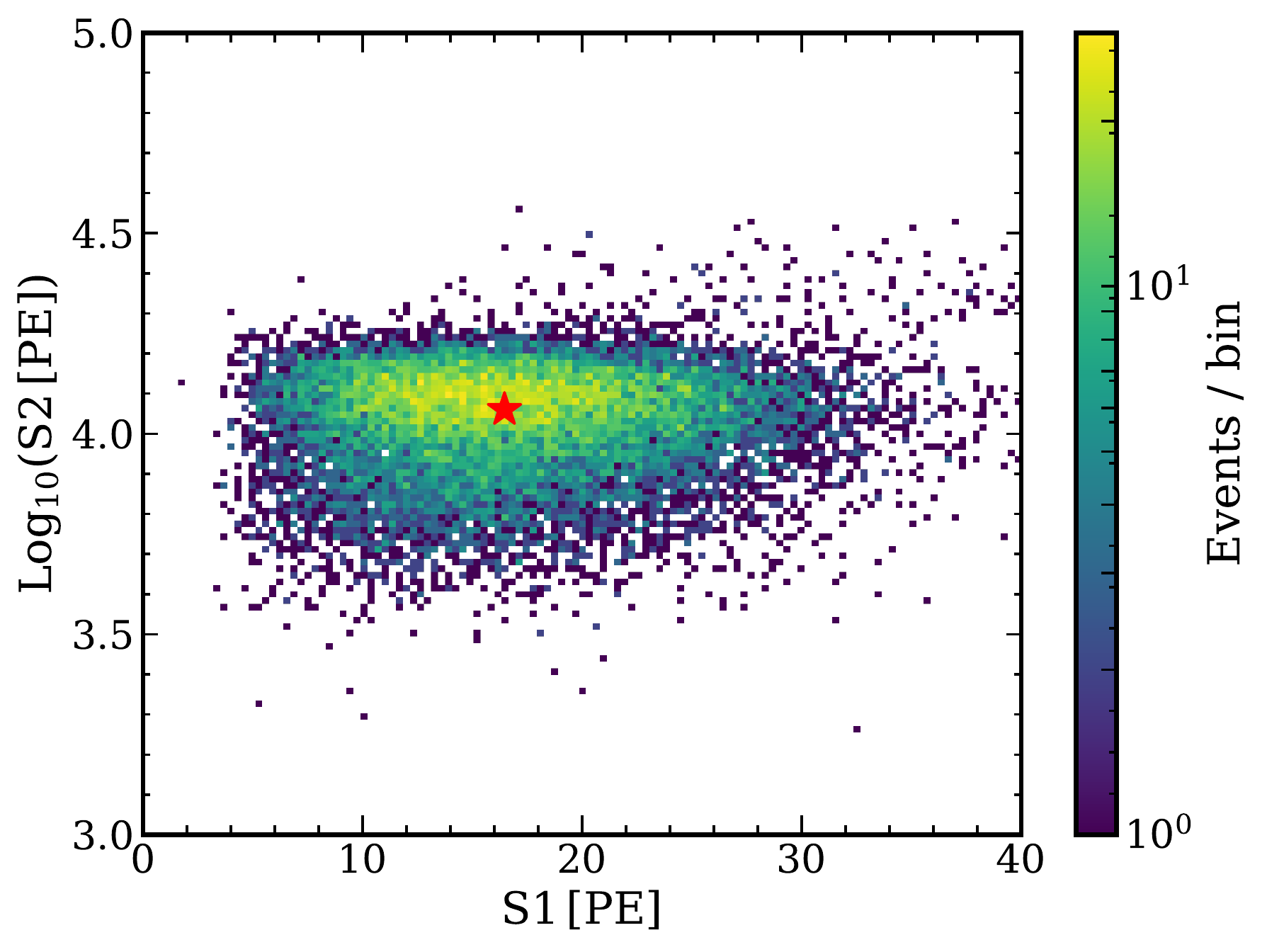}
    \end{minipage}\hfill
    \begin{minipage}{0.5\columnwidth}
        \includegraphics[width=\columnwidth]{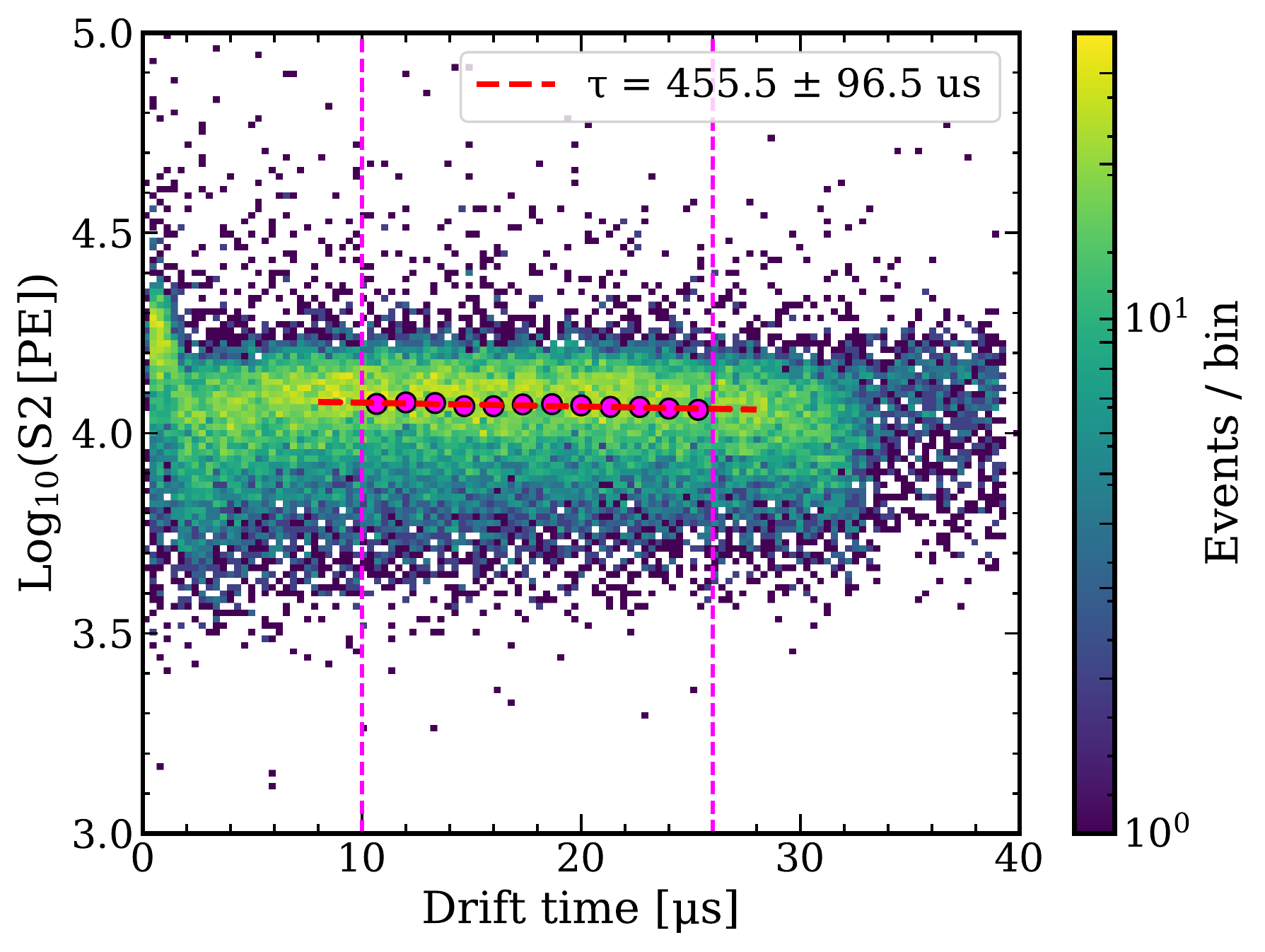}
    \end{minipage}\hfill
    \caption{(Left) Correlation of the observed S1 and S2 signals in number of photo-electrons (PE) from $^{57}$Co gamma rays interacting in the liquid xenon target.
             (Right) S2 as a function of drift time from these events. An exponential fit was applied to the mean S2 values (magenta dots) from each drift time bin in the range of [10, 26]\,$\mu s$ to obtain the electron lifetime (see text). The drift time window is selected to avoid the non-uniform drift field outside the range shown in figure\,\ref{extract_drift_field}. }
    \label{fig:co57_calibration}
\end{figure}
\FloatBarrier

A representative dataset is shown in figure\,\ref{fig:co57_calibration}.
Left plot shows the $\rm Log_{10}$(S2)\,-\,S1 space from $^{57}$Co. The location of the 122\,keV peak gives a mean S1 of 16.3 PE and S2 of 11,922 PE. The photon detection efficiency ($g1$) and the electron detection factor ($g2$) are obtained by comparing the observed S1 and S2 values with the yield of 122\,keV monoenergetic peak in the Noble Element Simulation Technique (NEST) calculator~\cite{nest}. Based on the NEST calculator, 5685 photons and 3344 electrons are expected from 122\,keV gamma line at 255\,V/cm drift field, giving $g1$\,=\,0.003\,$\rm PE/\gamma$ and $g2$\,=\,3.6\,$\rm PE/e^-$ in this detector setup. It is expected that this detector geometry would result in a much lower light collection efficiency, compared to the dark matter detectors such as XENON1T~\cite{Aprile:2017aty} and LUX~\cite{Akerib:2019psf}, due to the lack of a bottom PMT array and additional light transmission loss through the fused silica window. Nevertheless, the resulting $g2$ is sufficiently large, allowing us to observe single electrons drifting in the detector and to further study the single electron rate and its correlation with detector parameters. 

The S2 dependence on drift time\,($t_d$) is shown in the right of figure\,\ref{fig:co57_calibration}, where an exponential fit of S2($t_d$) = S2(0)$\cdot$exp(--$t_d$/$\tau$) was applied to derive the electron lifetime\,($\tau$). The magenta dots are the Gaussian mean of S2 in each $t_d$ bin. The middle part of the TPC, with a $t_d$ range of [10, 26]\,$\mu s$ and corresponding to the (255$\,\pm\,$8)\,V/cm field region in figure\,\ref{extract_drift_field} (left), was selected for the fitting, resulting in a measured electron lifetime\,($\tau$) of (456\,$\pm$\,96)\,$\mu s$ as indicated by the red dashed line in figure\,\ref{fig:co57_calibration} (right). The large uncertainty is caused by the limited drift time window and by operations on the circulation pipes, as described in Sec.\,\ref{sec:purification}. The S2 was slightly dropped with the drift time approaching to gate from $t_d$\,=\,$\sim$10\,$\mu s$, and with drift time approaching to the TPC bottom from $t_d$\,=\,$\sim$26\,$\mu s$, caused by the field deformation mentioned in Sec.\,\ref{sec:tpcdesign}. The data with drift time below 10\,$\mu s$ and above 26\,$\mu s$ was not included in the electron lifetime fitting.


\subsection{Time evolution of purification}
\label{sec:purification}

\begin{figure}[hbt!]
\centering
	\begin{minipage}{0.9\columnwidth}
		\centering
		\includegraphics[width=\columnwidth]{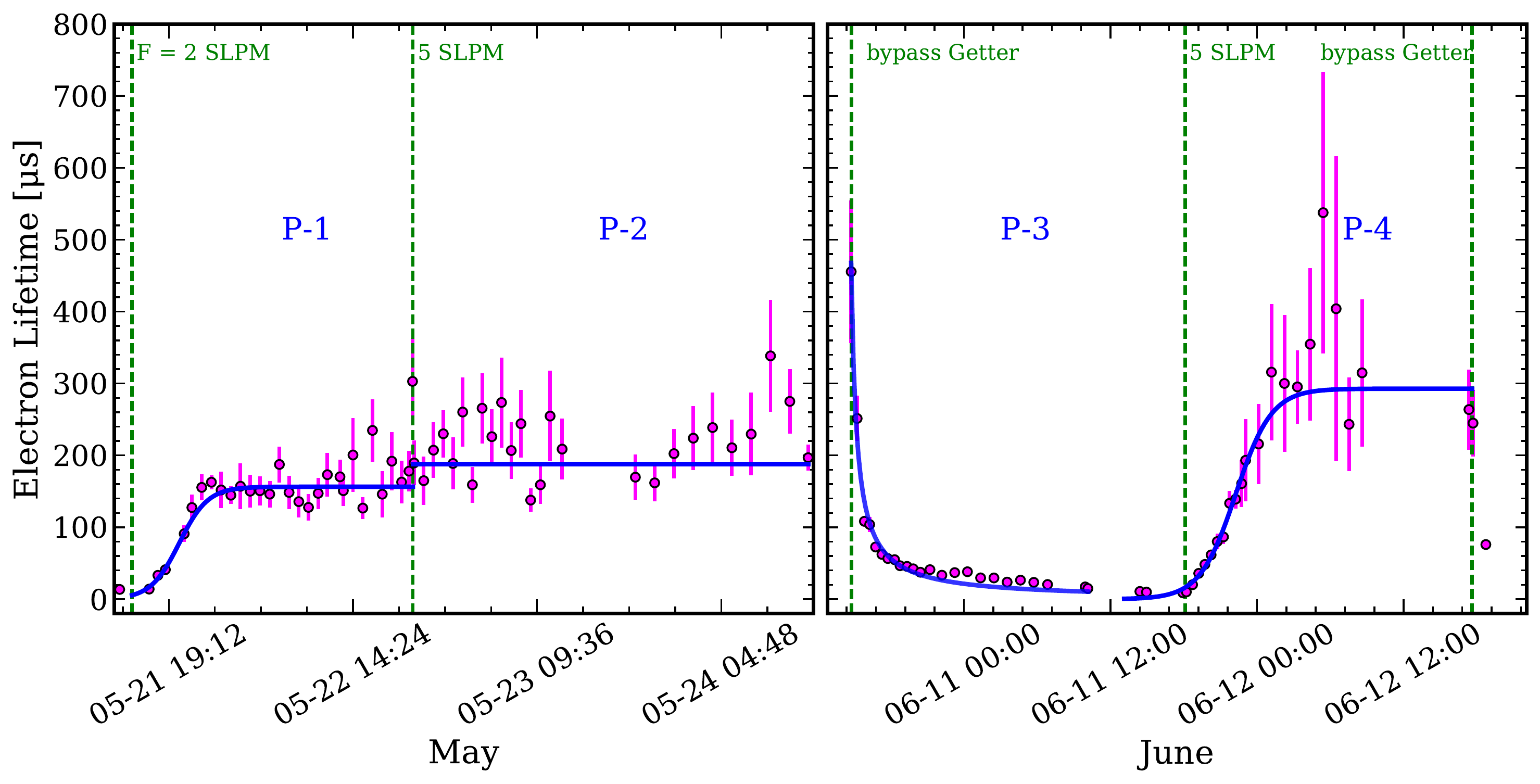}
	\end{minipage}
\caption{Electron lifetime as a function of time for different purification settings. A simple purification model (blue lines, see text) is used to fit the data to obtain parameters relevant to the out-gassing and liquid purity. The highest electron lifetime obtained is a little over 0.5 ms, with large uncertainties due to the limited drift time window and operations on circulation pipes upgrade. Data were taken in 2019.}
\label{fig:eltred}
\end{figure}

To investigate the purification efficiency of the sealed TPC, we monitor the electron lifetime at different gas xenon circulation and purification speeds. The getter used (SAES model PS3-MT3-R-2) has a maximum purification speed of 5\,standard-liter-per-minute\,(SLPM) for nitrogen, helium and argon, but 3\,SLPM for xenon according to the specifications. During the detector operation, the xenon flow rate through the getter was set at different values to investigate the purification efficiency and understand the xenon purity. The time evolution of electron lifetime $\tau$ was monitored for several days using the calibration data and divided into four periods as shown in figure\,\ref{fig:eltred}. 

\FloatBarrier

A simple purification model was developed to describe the electron lifetime evolution during our run. During the gas xenon circulation and purification, the impurity concentration in the target LXe can be described as,

\begin{equation}
    \frac{M}{\rho}\frac{dn\left(t\right)}{dt} = R_o - n\left(t\right)\eta F
    \label{eq1}
\end{equation}

where $n\left(t\right)$ is the oxygen-equivalent electro-negative impurity concentration in unit of part-per-billion (ppb) in xenon. $M$ is the total mass of LXe inside the TPC ($\sim$0.54\,kg in this design). $\rho$ is the density of xenon gas (5.9\,$\times 10^{-3}$\,kg/liter). $R_o$ is a constant representing the \textit{effective} out-gassing rate of impurities released into the active LXe target. Note that the out-gassing here is not the same concept in vacuum. It includes both the out-gassing from detector materials, impurities dissolved from material surfaces contacting LXe, and impurities leaked from the outside of the sealed TPC into the active LXe target. $F$ is the xenon gas circulation flow rate. We added the $\eta$ term to describe the purification efficiency in case the impurity in the xenon is not fully removed by the getter, or if additional impurities are added to the purified xenon gas before it reaches the target volume inside the TPC.

The time evolution of the electron lifetime, at a constant purification speed $F$, thus can be obtained as,

\begin{equation}
    \tau\left(t\right) = \frac{368}{n_0e^{-\frac{\rho\eta F}{M}t} + \frac{R_o}{\eta F}\left(1-e^{-\frac{\rho\eta F}{M}t}\right)}\;\left[\mu s\right]
    \label{eq2}
\end{equation}

where $n_0$ is the initial (t\,=\,0) concentration. The constant 368 is the electric field dependent term which can be derived from Ref.\,\cite{Bakale:1976} to relate the electron lifetime and impurity concentration in LXe as, $\tau\left(t\right) \approx \frac{368}{n\left(t\right)} [\mu s]$. After reaching equilibrium, the electron lifetime is expected to reach a plateau of

\begin{equation}
    \tau\left(t\to\infty\right) = \frac{368\times\eta F}{R_o}\;\left[\mu s\right]
    \label{eq3}
\end{equation}

Each of the four periods in figure\,\ref{fig:eltred} is fitted with the relevant purification model equations. The fit results are summarized in Tab.\,\ref{tab:model}. The period of P-1 is fitted with Eq.\,\ref{eq2} with $F$ = 2\,SLPM while the P-2 is fitted with a constant value of Eq.\,\ref{eq3} suppose the maximum electron lifetime has been reached after the flow rate being switched to $F$ = 5\,SLPM. The model gives 0.51$\pm$0.02 and 0.25$\pm$0.01 of $\eta$ for P-1 and P-2, respectively. Less than 100\% purification efficiency is obtained in P-1 due to the highly contaminated LXe outside of the sealed TPC compared to the cleaner active LXe in the sealed TPC. The further drop of purification efficiency from P-1 to P-2 indicates the degradation of the getter's purification capability after reaching its maximum specified flow rate (3 SLPM) for xenon. The effective impurity out-gassing rate $R_o$ of $(4.0\pm0.3)\times 10^{-11}$~liters/s is obtained from the fit. 

The period of P-4 is fitted with Eq.\,\ref{eq2}. A $\eta$ of (0.21\,$\pm$\,0.02) is achieved which is consistent with P-2 at the same flow rate of 5\,SLPM. Note that the large fluctuation of $\tau$ in June is mostly caused by the operations on circulation pipes for the purification system upgrade and limited drift time window in electron lifetime fitting. A higher electron lifetime was achieved in June compared to May, which indicates that the out-gassing outside the TPC is smaller in June compared to May after $\sim$1 month of purification, which resulted in a lower effective impurity out-gassing rate $R_o$ of $(2.0\pm0.3)\times 10^{-11}$~liters/s.

\begin{table}[htbp]
\centering
\caption{\label{tab:model}The fit results from figure\,\ref{fig:eltred}. Note that the $R_o$ from P-1 fitting is used in Eq.\,\ref{eq3} for  P-2 fitting.}
\bigskip
\begin{tabular}{|c|c|c|c|c|c|}
\hline
Periods   &  Circulation & $n_0$\,[ppb]  &  $R_o$\,[10$^{-11}$\,liter/s]  &  $\eta$  &  $n_{out}$\,[ppb]  \\
\hline
P-1       &  2 SLPM through getter & 66.2 $\pm$ 7.0    &  4.0 $\pm$ 0.3  &  0.51 $\pm$ 0.02   &   n/a                     \\
P-2       &  5 SLPM through getter &   n/a            &      n/a           &  0.25 $\pm$ 0.01   &   n/a                     \\
P-3       &  5 SLPM bypass getter & 0.79 $\pm$ 0.08   &   negligible       &  n/a               &   32.4 $\pm$ 2.4          \\
P-4       & 5 SLPM through getter & 22.5 $\pm$ 3.0    &  2.0 $\pm$ 0.3   &  0.21 $\pm$ 0.02   &   n/a                    \\
\hline
\end{tabular}
\end{table}

During the period of P-3, the getter was bypassed for understanding the purification process, but the flow was kept at 5\,SLPM in the bypass pipes; thus, the impurity concentration in LXe can be described by Eq.\,\ref{eq4}, resulting in \ref{eq5}.

\begin{equation}
    \frac{M}{\rho}\frac{dn\left(t\right)}{dt} = R_{o} + n_{out}F
    \label{eq4}
\end{equation}

\begin{equation}
    \tau\left(t\right) = \frac{368}{n_0 + \frac{\left(R_o + n_{out}F\right)\rho}{M}t}\;\left[\mu s\right]
    \label{eq5}
\end{equation}

where $n_{out}$ is the impurity concentration in the LXe outside the sealed TPC where the xenon is taken out through the circulation. In this diagnostic run, impurities from outside of the sealed TPC are brought inside quickly, reducing the liquid xenon purity and thus lowering the electron lifetime. 

By fitting the period of P-3 with Eq.\,\ref{eq5}, (2.7\,$\pm$\,0.2)\,$\times$\,$10^{-9}$\,liter/s of ($R_o$\,+\,$n_{out}F$) was obtained. The $R_o$ is very small comparing to the values obtained from P-1  and P-4 periods. An impurity concentration of (32.4\,$\pm$\,2.4)\,ppb is obtained in the liquid xenon outside the sealed TPC, considering the 5\,SLPM flow rate $F$ and negligible $R_o$. This indicates the LXe outside the sealed TPC contains a much higher concentration of impurities than that inside. In the current design of the sealed TPC, the LXe in the sensitive target is always first mixed with the LXe outside before passing through the getter, effectively increasing the burden of the getter to purify the LXe outside the target and reducing the purification efficiency. In addition, the impurity gas molecules from the outside the sealed TPC can \textit{leak} into the active LXe target through the small slit between the anode and gate, increasing the value of $R_o$.  A LXe detector with the target volume completely sealed will not only have reduced $R_o$ but also improved purification efficiency; thus, the electron lifetime can be further improved. A completely sealed TPC would require a different cryogenic and purification system design and will be pursued in the future. 


\subsection{Single electron detection}
\label{sec:low_s2}

Understanding and controlling the background at the single- to few-electron level is critical to improving the sensitivity to sub-GeV dark matter electron scattering and to CE$\nu$NS of low energy reactor neutrinos. There are two types of single/few electrons background. One is the \textit{prompt} single electron background within one max drift time window, typically at tens or hundreds $\mu$s following a large signal (S1 or S2). The other type is the \textit{delayed} single electrons that can extend to hundreds of ms following a large signal (S1 or S2). In this study, we investigate the  \textit{prompt} single electron rate and its correlation with impurity levels in LXe or different electrodes. The study of \textit{delayed} single electron emission~\cite{Akerib:2020jud} requires much lower background and will be pursued in the future.

\subsubsection{Single electron identification}
\label{sec:e_obs}

\begin{figure}[hbt!]
\centering
\begin{minipage}{0.5\columnwidth}
\centering
\includegraphics[width=\columnwidth]{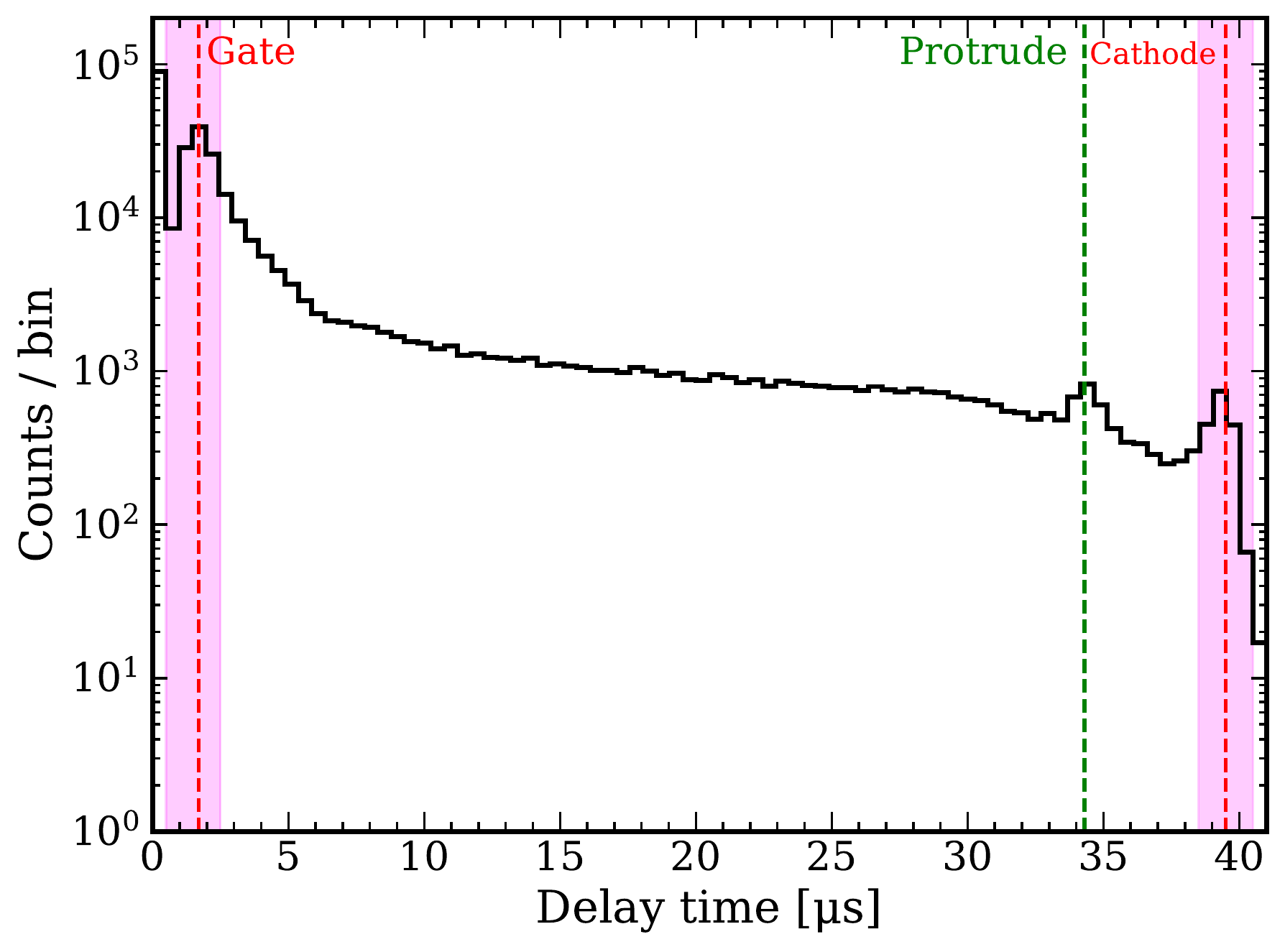}
\end{minipage}
\caption{Delay time of the small S2s, corresponding to single or a few electrons, after the main S2 from a recoil event. The red dashed lines indicate the location of the gate and cathode electrodes respectively, while the green dashed line is the location of the protruding edge as shown in figure\,\ref{fig:TPC_sketch}\,(right).}
\label{single_e_obs}
\end{figure}

The raw data for the single electron study was acquired with window length of 80\,$\mu s$ with 50\% post trigger. Most of the events were triggered on an S2 peak; thus $\sim$40\,$\mu s$ time window after S2 can be used for extracting the single electrons from the photoionization by S2 light. Figure\,\ref{single_e_obs} shows the delay time of the small S2s after the main one. Three clear peaks can be observed, indicating the location of the gate, protruding edge and cathode respectively. The peaks (red dashed lines) around gate and cathode electrodes are considered to be from the photo-electron emission from electrode surface by the main S2. The peak (green dashed line) around the protruding edge is most likely caused by the electrons accumulation on the edge of the acrylic material. The other region is mainly the photoionization of the main S2 light on the impurities in the bulk LXe.

\FloatBarrier

\begin{figure}[ht]
    \centering
    \begin{minipage}{0.49\columnwidth}
        \includegraphics[width=\columnwidth]{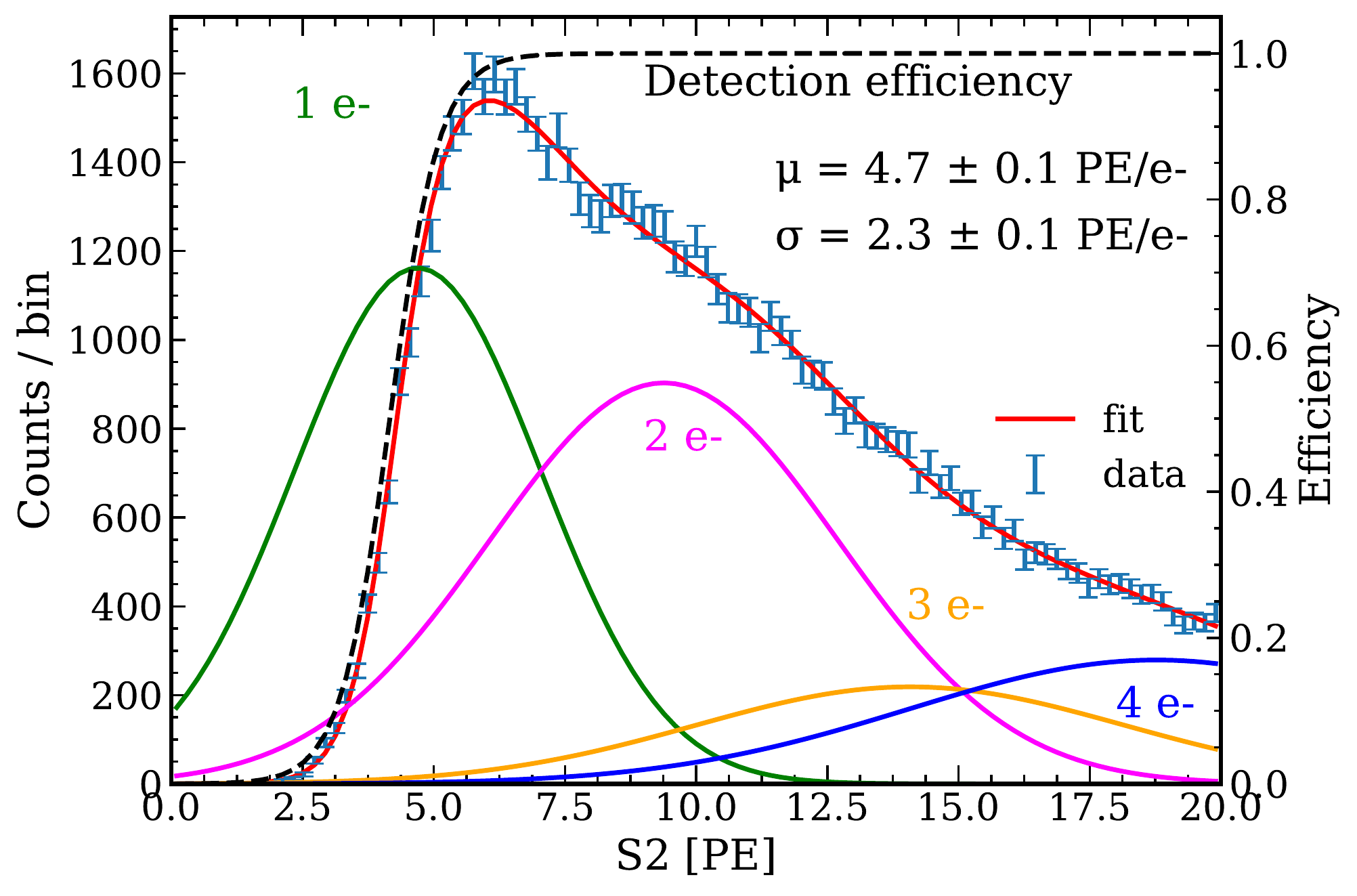}
    \end{minipage}\hfill
    \begin{minipage}{0.49\columnwidth}
        \includegraphics[width=\columnwidth]{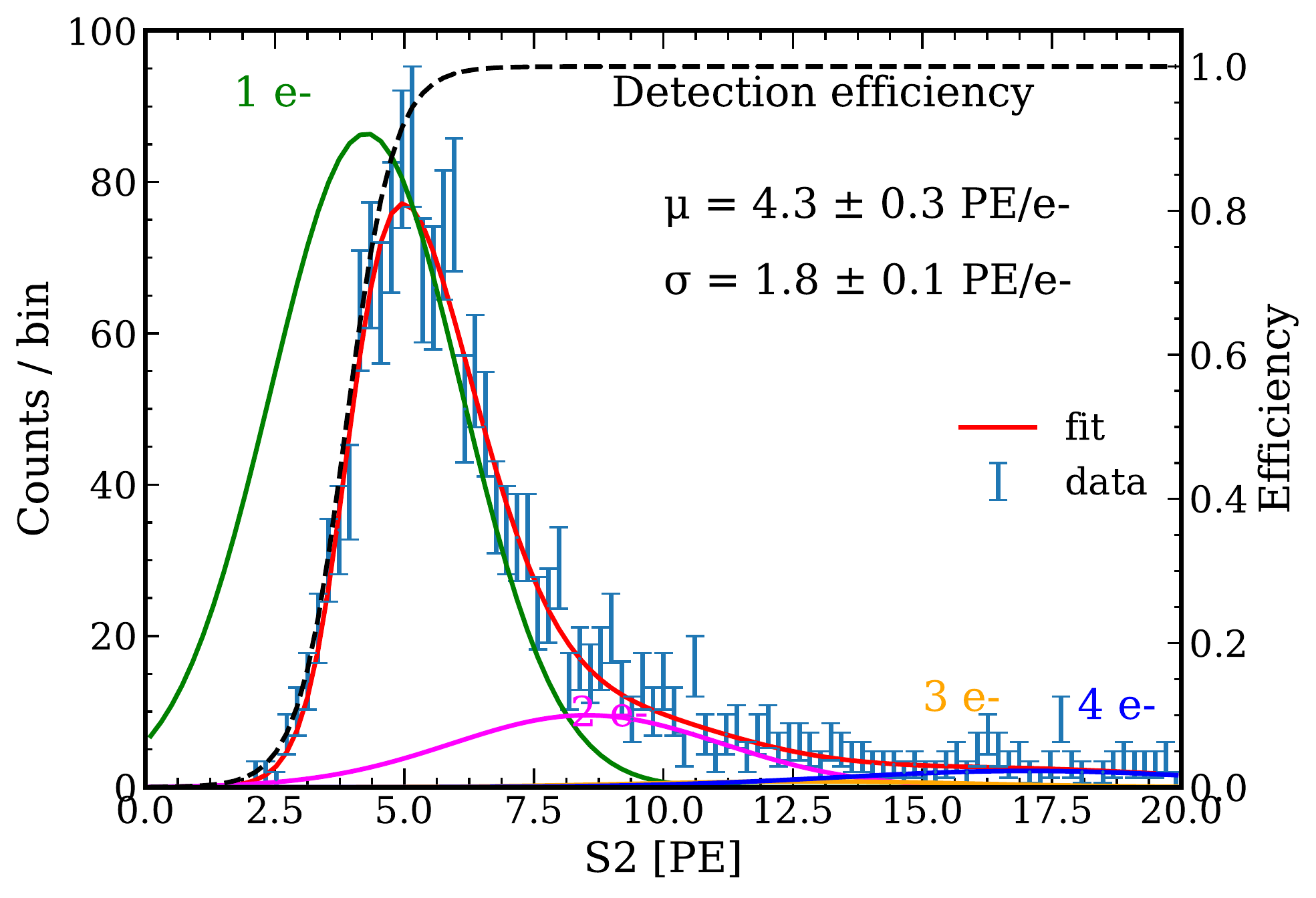}
    \end{minipage}
    \caption{Typical small S2s spectrum within 2\,$\mu s$ time window for gate (left) and cathode (right) electrode respectively.  The spectrum was fitted with a sum of 4 Gaussians, supposing that the spectrum comprises a sum of one to four electrons, multiplied by a function (see text) to take into account the detection efficiency.}
    \label{single_e_fit}
\end{figure}
\FloatBarrier

To extract the single electron signals, the events within 2\,$\mu s$ time window around the peak of the gate (or cathode), as shown in the pink shaded region in figure\,\ref{single_e_obs}, were selected. The corresponding S2 spectra are shown in figure\,\ref{single_e_fit}. 
The spectra are fitted with a sum of several Gaussian functions multiplied by an efficiency curve which is expressed as Eq.\,\ref{sefit}. The efficiency curve is described by an empirical function of $\frac{1}{e^{-\left(S2 - A\right)/B} + 1}$ with $A$ and $B$ as free parameters. The Gaussian functions consists of one Gaussian with mean $\mu$ and standard deviation $\sigma$ to describe the single electron component, followed by a few electrons contributions following Poissonian statistics and their Gaussian parameters are thus dependent on the single electron values.

\begin{equation}
    f(S2) = \frac{1}{e^{-\frac{S2 - A}{B} + 1}} \times \sum_{i=1}^{4}A_{i}e^{-\frac{(S2-i\mu)^2}{2i\sigma^2}}
    \label{sefit}
\end{equation}

This fit assumes that the S2 spectrum comprises a sum of one to a few electrons S2 signals with the first Gaussian ($\mu$, $\sigma$) provides the detected PEs per electron which has been extracted to GXe. The efficiency curve reflects the efficiency of the S2 peak-finder algorithm. 
The single electron gain ($\mu\pm\sigma$) of 4.7\,$\pm$\,2.3\,PE/$\rm e^-$ and 4.3\,$\pm$\,1.8\,PE/$\rm e^-$ were derived from the data for gate and cathode electrodes respectively, which is consistent with each other. In addition, the single electron gain can also be estimated based on the $g2$ and electrons extraction efficiency at gas-liquid interface. The extraction field is 5.4\,kV/cm in LXe from Comsol simulation corresponds to a extraction efficiency of $\sim$90\%~\cite{Xu:2019dqb}. Considering $g2$\,=\,3.6\,$\rm PE/e^-$ from Sec.\,\ref{sec:co_cali}, the single electron gain is 4.0\,PE/$\rm e^-$, which is consistent with the fit results. 

\subsubsection{Photoionization on electrodes and LXe impurities}
\label{sec:impurity}

\begin{figure}[ht]
    \centering
    \begin{minipage}{0.48\columnwidth}
        \includegraphics[width=\columnwidth]{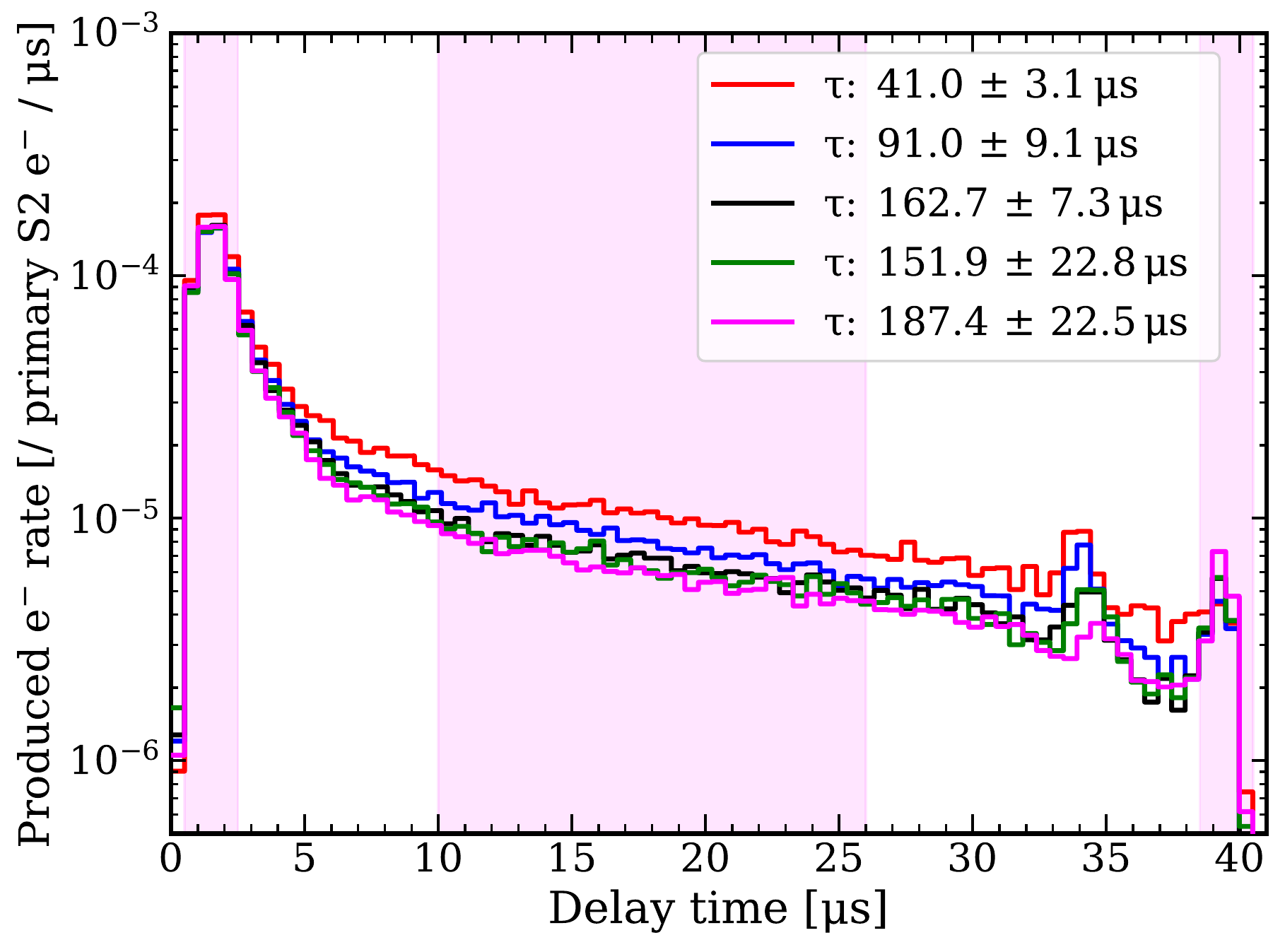}
    \end{minipage}\hfill
    \begin{minipage}{0.48\columnwidth}
        \includegraphics[width=\columnwidth]{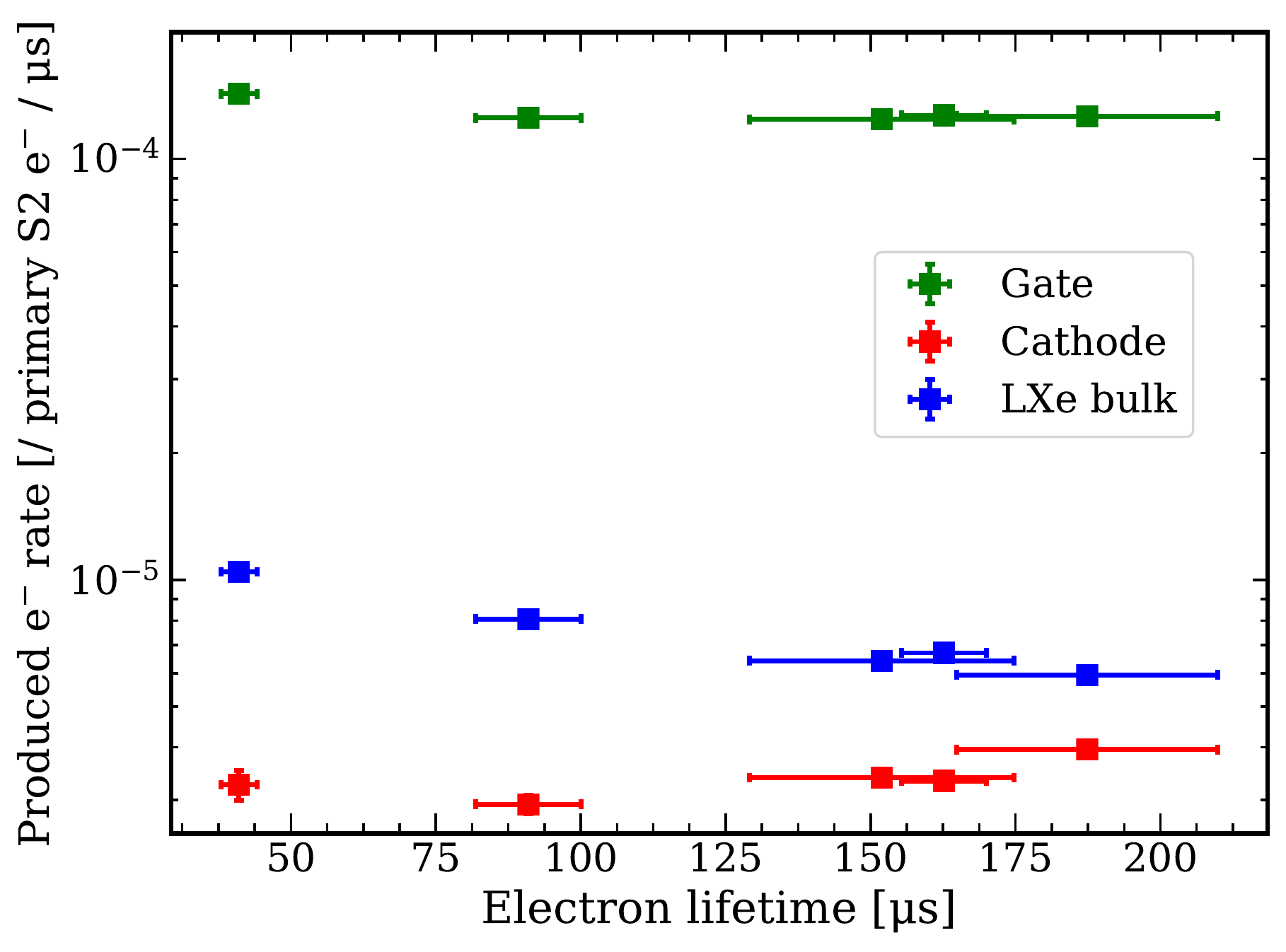}
    \end{minipage}
    \caption{(Left) The single electron rate as a function of the delay time at different electron lifetime. see more explain in text.
             (Right) The single electron rate from gate, cathode and LXe bulk region, as a function of its electron lifetime. The rate is the average value from the corresponding shaded-band in the left plot.}
    \label{single_e_source}
\end{figure}

To understand the photoionization mechanisms from different materials, the single electron rate from the gate, cathode and LXe bulk are investigated. The single electron rate is defined as the number of produced electrons per $\mu$s by the S2 light from a primary electron drifting in the gas phase. The single electron rate as a function of its delay time is shown in figure\,\ref{single_e_source} (left) for runs with different electron lifetime (or liquid purity). The rate was corrected for the loss of detected electrons due to impurity in the liquid, thus the rate plotted in figure\,\ref{single_e_source} (left) is the actual single electron rate produced either on the electrode or at the exact location in the liquid. 

For electrons generated on the gate and cathode electrodes, we used the average rate within 2\,$\mu s$ delay time window around the peak rate. For electrons generated on the impurities in the bulk LXe, we chose the rate within [10, 26]\,$\mu s$ delay time window in the center of the TPC to avoid the potential field distortion in the top and bottom regions. The corresponding delay time regions are the shaded-bands in figure\,\ref{single_e_source} (left). The rates from each component are shown in figure\,\ref{single_e_source} (right). The single electron rate from LXe bulk decreases with the increasing of electron lifetime due to the purity improvement, while the rates from the gate and cathode are less affected by the LXe impurities as expected.

\FloatBarrier

\begin{figure}[ht]
    \centering
    \begin{minipage}{0.48\columnwidth}
        \includegraphics[width=\columnwidth]{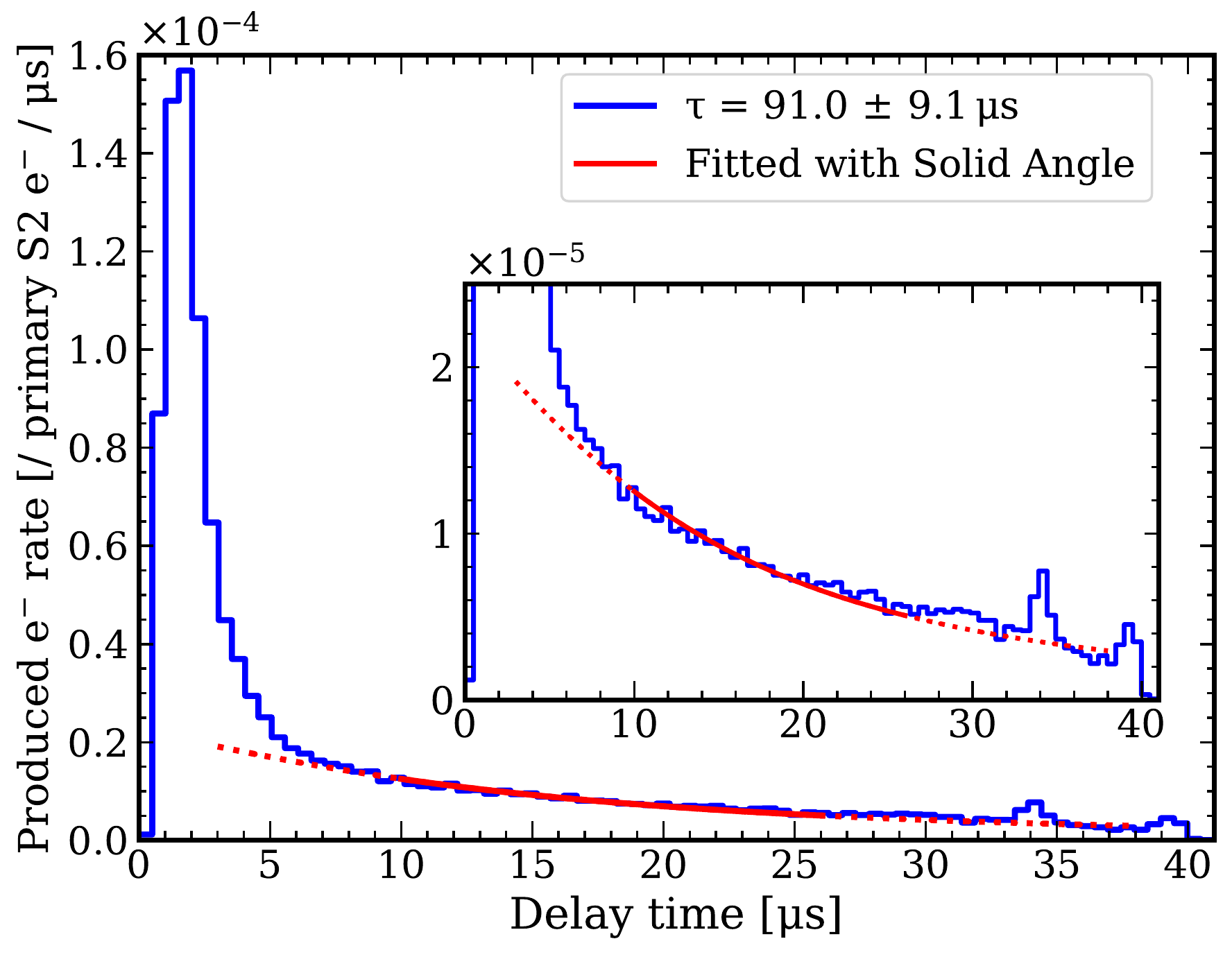}
    \end{minipage}\hfill
    \begin{minipage}{0.48\columnwidth}
        \includegraphics[width=\columnwidth]{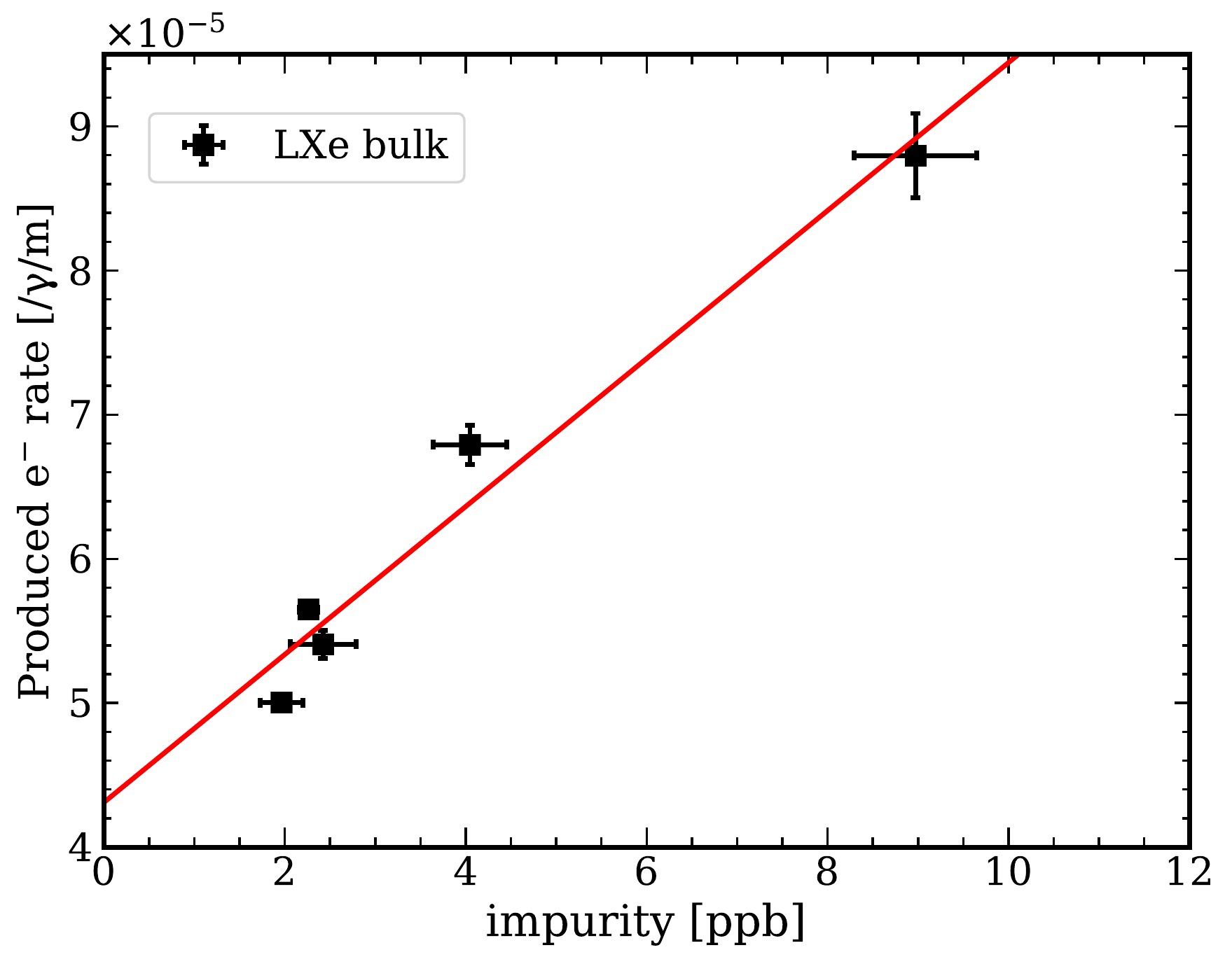}
    \end{minipage}
    \caption{(Left) The single electron rate as a function of its delay time for data with 91.1\,$\mu s$ electron lifetime, fitted with the solid angle model in range of [10, 26]\,$\mu s$ delay time. The inset is a zooming of the fit region.
            (Right) Photoionization rate as a function of the impurities in the LXe bulk. The impurity in $\rm ppb$ is calculated from the electron lifetime (see text for details). The data-points is fitted with a linear function.}
    \label{single_e_impurity}
\end{figure}

Although the single electron rate has been corrected with electron lifetime, it still decreases with the delay time. This can be explained by the smaller number of S2 photons hitting the lower part of the TPC. A simple Solid Angle\,(SA) model can be used to model the photon emission from the location where S2 is produced. The S2 photons are mostly produced around anode wires due to the strong local electric field, and emitted isotropically. In the SA model, we assume the S2 is produced in the center of the anode electrode, thus the solid angle inside the TPC gradually decreases with the increasing of TPC's depth. One example single electron rate as a function of its delay time is shown in figure\,\ref{single_e_impurity} (left). The rate between the delay time of [10, 26]\,$\mu s$ was fitted with the SA model. The fit is slightly deviated from the data caused by the photon reflection on the thin PTFE sheet on TPC wall. This distribution indicates that the single electrons from the LXe bulk are most likely associated with the neutral impurities\,\cite{Akerib:2020jud} which are more evenly distributed in LXe than negative ions caused by electrons attached to the impurities. The larger deviation from the SA model for events near the gate electrode is most likely caused by the inaccurate drift time correction due to the field distortion as described in Sec.\,\ref{sec:co_cali}.

\FloatBarrier

The photo-electron emission or photoionization probability can be estimated for different electrode materials and different LXe impurities level as following. The number of photons that is produced by one extracted electron in GXe is calculated by the empirical equation: $N_{\gamma} = \alpha(\varepsilon_e/p - \beta)px$. $\alpha$ is the amplification factor and $\beta$ is the threshold of reduced field for proportional light production. $\alpha$ = 120, and $\beta$ = 1.3 kV/cm/atm\,\cite{Aprile:2004kx} are used in our calculation. The electric field $\varepsilon_e$ of 9.6\,kV/cm in S2 production region is from the COMSOL simulation. In our measurement, the gas pressure ($p$) is controlled and kept at 2.2 atm. The gap ($x$) for proportional light production is 0.25\,cm. In the end, 202\,$\gamma$/$\rm e^-$ of  $N_\gamma$ can be derived based on the parameters above. 

To estimate the number of photons hitting the electrodes, a simple Geant4 simulation was performed. A full geometry of the sealed TPC was implemented except the mesh electrodes. Assuming a conservative PTFE reflectivity of 95\% in the simulation, a light collection efficiency ($LCE$) of 43.5\% and 31.3\% was achieved for the location at the gate and cathode respectively. Considering the gate transparency of 92\%, the number of photons hitting the gate can be estimated as 202\,$\times$\,43.5\%\,$\times$\,(1-92\%). The number of electrons emitted from the gate can be calculated by integrating 2\,$\mu s$ time window around gate in the histogram of figure\,\ref{single_e_source} (left). Based on the number of incident photons and emitted electrons, the probability of photo-electron emission of 7.4$\,\times10^{-5}$\,$e^-/\gamma$ can be obtained for the stainless steel surface, which is in consistent with the value of $\sim$1.0\,$\times10^{-4}$\,$e^-/\gamma$ in Ref.\,\cite{Laulainen_2015}. For the graphene cathode, assuming 96\% light transparency of the single-layer graphene\,\cite{Bae_2010}, the probability of photo-electron emission of graphene from LXe scintillation light is estimated to be 5.8\,$\times10^{-6}$\,$e^{-}$\,/$\gamma$, which is one order of magnitude lower than that of stainless steel. While the work function of stainless steel (4.3\,eV\,\cite{1966JAP....37.2261W}) is lower than graphene (4.5$\sim$4.8\,eV\,\cite{Yu_2009}), there could be other reasons such as the thicker surface of the stainless steel that contributes to a larger photo-electron emission probability. 

In LXe, the single electron rate can be explained by the photoionization on oxygen-equivalent impurities. The cross section of the xenon scintillation light on oxygen atoms is $\sim$1\,Mb with very low efficiency (<1\%)\,\cite{Brion:1979}. For 1\,ppb concentration of oxygen in LXe, it gives a  photoionization rate of $\sim$1.4\,$\times10^{-5}$\,$e^{-}$\,/$\gamma$\,/$m$. In our measurement, the photoionized electrons within [10, 26]\,$\mu s$ delay time from each extracted electron can be calculated by integrating the histogram in figure\,\ref{single_e_source}\,(left). The number of photons cross this LXe region can be estimated using the Geant4 simulation. Assuming a 2.5-5.0\,cm mean path length for photons within the [10, 26]\,$\mu s$ delay time window, the photoionization rate in LXe was calculated and shown in figure\,\ref{single_e_impurity}\,(right) for different electron lifetimes. A linear fit results in a value of (3-5)\,$\times 10^{-6}$\,$e^{-}$\,/$\gamma$\,/$m$ for every ppb impurity concentration, which is compatible with the calculation from the photoionization cross section on oxygen but lower than the roughly estimated value of (5-20)\,$\times10^{-5}$\,$e^{-}$\,/$\gamma$\,/$m$ from the LUX experiment\,\cite{Akerib:2020jud}.


\section{Conclusion}
\label{sec:concl}

In this work, we presented a sealed LXeTPC with a graphene-coated fused silica window as the cathode electrode. The effectiveness of the LXe purification is studied using the electron lifetime evolution data with a simple purification model. It shows a much cleaner liquid xenon purity in the LXe target inside the sealed chamber, compared to that outside, achieving an electron lifetime of 0.5 ms with uncertainties limited by the drift length of the TPC. We expect further improvement of the purification efficiency with a completely-sealed TPC, which will be developed in the future. 

We investigated the single electron rate caused by the photoelectric effect on the electrode materials and the photoionization on impurities in LXe. The photoelectron emission probability of graphene is found to be lower than that of stainless steel. The photoionization rate of impurities in LXe shows that the \textit{prompt} single electrons are caused mainly by the oxygen impurities in the bulk LXe. 

The TPC concept demonstrated here serves as a basis for further development of the sealed TPC technique, with the goal to significantly improve the purification efficiency of LXe in the target and to reduce the single/few electrons background for future experiments to search for sub-GeV dark matter via electron scattering~\cite{Bernstein:2020cpc}, reactor neutrino detection via the CE$\nu$NS process~\cite{Akimov:2019ogx}, and the next generation multi-purpose liquid xenon experiment, e.g. DARWIN~\cite{Aalbers:2016jon}, for dark matter and neutrino physics. 

\acknowledgments

This material is based upon work supported by the U.S. Department of Energy, Office of Science, Office of High Energy Physics under Award Number \text{DE-SC0018952}.


\bibliographystyle{unsrt}
\bibliography{bibliography}

\end{document}